\documentclass[a4paper,12pt]{article}
\usepackage[dvipdfmx]{graphicx}
\usepackage{amsfonts,amsthm,amsmath,amssymb,graphicx,float,mathtools}
\usepackage{xcolor}
\usepackage{ulem}
\numberwithin{equation}{section}
\usepackage{subfig}
\usepackage{color}
\usepackage{appendix}
\usepackage{bbm} 
\usepackage{tensor} 
\usepackage{verbatim} 
\usepackage[
	  pagebackref=false,
	  colorlinks=true,
      linkcolor=blue,
      urlcolor=blue,
      filecolor=black,
      citecolor=red,
      pdfstartview=FitV,
      pdftitle={},
        pdfauthor={},
        pdfsubject={},
        pdfkeywords={},
        pdfpagemode=None,
        bookmarksopen=true
      ]{hyperref}
\usepackage{ascmac}
\graphicspath{ {Pic/} }

\begin{document}

\newtheorem{definition}{Definition}[section]
\newcommand{\be}{\begin{equation}}
\newcommand{\ee}{\end{equation}}
\newcommand{\bea}{\begin{eqnarray}}
\newcommand{\eea}{\end{eqnarray}}
\newcommand{\LE}{\left[}
\newcommand{\R}{\right]}
\newcommand{\nn}{\nonumber}
\newcommand{\Tr}{\text{Tr}}
\newcommand{\N}{\mathcal{N}}
\newcommand{\G}{\Gamma}
\newcommand{\vf}{\varphi}
\newcommand{\LL}{\mathcal{L}}
\newcommand{\Op}{\mathcal{O}}
\newcommand{\HH}{\mathcal{H}}
\newcommand{\arctanh}{\text{arctanh}}
\newcommand{\up}{\uparrow}
\newcommand{\down}{\downarrow}
\newcommand{\ket}[1]{\left| #1 \right>}
\newcommand{\bra}[1]{\left< #1 \right|}
\newcommand{\ketbra}[1]{\left|#1\right>\left<#1\right|}
\newcommand{\rd}{\partial}
\newcommand{\de}{\partial}
\newcommand{\ba}{\begin{eqnarray}}
\newcommand{\ea}{\end{eqnarray}}
\newcommand{\db}{\bar{\partial}}
\newcommand{\we}{\wedge}
\newcommand{\ca}{\mathcal}
\newcommand{\lr}{\leftrightarrow}
\newcommand{\f}{\frac}
\newcommand{\s}{\sqrt}
\newcommand{\vp}{\varphi}
\newcommand{\hvp}{\hat{\varphi}}
\newcommand{\tvp}{\tilde{\varphi}}
\newcommand{\tp}{\tilde{\phi}}
\newcommand{\ti}{\tilde}
\newcommand{\ap}{\alpha}
\newcommand{\pr}{\propto}
\newcommand{\mb}{\mathbf}
\newcommand{\ddd}{\cdot\cdot\cdot}
\newcommand{\no}{\nonumber \\}
\newcommand{\la}{\langle}
\newcommand{\lb}{\rangle}
\newcommand{\ep}{\epsilon}
\renewcommand{\thesubfigure}{\Roman{subfigure}}
 \def\we{\wedge}
 \def\lr{\leftrightarrow}
 \def\f {\frac}
 \def\ti{\tilde}
 \def\ap{\alpha}
 \def\pr{\propto}
 \def\mb{\mathbf}
 \def\ddd{\cdot\cdot\cdot}
 \def\no{\nonumber \\}
 \def\la{\langle}
 \def\lb{\rangle}
 \def\ep{\epsilon}
\newcommand{\mcl}{\mathcal}
 \def\g{\gamma}
\def\Tr{\text{tr}}

\begin{titlepage}
\thispagestyle{empty}

\begin{flushright}

\end{flushright}
\bigskip

\begin{center}
  \noindent{\large \textbf{Time Ordering Effects and Destruction of Quasiparticles in Two-dimensional Holographic CFTs }}
\vspace{2cm}

\vspace{1cm}
\renewcommand\thefootnote{\mbox{$\fnsymbol{footnote}$}}
Weibo Mao \footnote{maoweibo21@mails.ucas.ac.cn}${}^{1}$, and 
Masahiro Nozaki\footnote{mnozaki@ucas.ac.cn}${}^{1,2}$\\

\vspace{1cm}
${}^{1}${\small \sl Kavli Institute for Theoretical Sciences, University of Chinese Academy of Sciences,
Beijing 100190, China}\\
${}^{2}${\small \sl RIKEN Interdisciplinary Theoretical and Mathematical Sciences (iTHEMS), \\Wako, Saitama 351-0198, Japan}\\

\vskip 4em
\end{center}
\begin{abstract}
In this paper, we investigate the entanglement dynamics induced by a composite operator, defined as the local operator evolved with the time evolution operator constructed of the Euclidean and Lorentzian ones.
The systems under consideration are described by two-dimensional holographic conformal field theories ($2$d holographic CFTs), the theories described by gravity.
Then, we find that the time ordering of Euclidean and Lorentzian time evolutions affects the behavior of bipartite entanglement and non-local correlation, especially their late-time ones.
We also establish the relation between entanglement entropy and energy-momentum densities.
This relation states that the late-time behaviors of bipartite entanglement and non-local correlation strongly depend on the energy distribution. We also examine how the decay of the quasiparticle influences the bipartite entanglement and non-local correlation.
Furthermore, we investigate the gravity dual of the systems considered in this paper.
\end{abstract}
\end{titlepage} 
\tableofcontents
\tableofcontents

\section{Introduction}
Over the past few years, it has been one of the central research subjects in holography, the equivalence between the quantum field theories (QFTs) and the theories containing the gravity \cite{1999IJTP...38.1113M}, to investigate the dynamical feature of the holographic QFTs, the quantum field theories possessing their gravity duals.
An expected feature of the holographic QFTs is strong scrambling ability \cite{2007JHEP...09..120H,2008JHEP...10..065S,2013JHEP...04..022L,2013JHEP...05..014H,2014JHEP...12..046S,2014JHEP...03..067S,Cotler:2016fpe,2016JHEP...08..106M,2016JHEP...02..004H}. This feature may induce the growth of operator \cite{2017PhRvX...7c1016N,2018PhRvX...8b1013V,2018JHEP...06..122R} and quantum thermalization (the thermalization of the subsystem) \cite{Liu:2013qca,PhysRevX.7.031016}. 
We further mention the breakdown of the quasiparticle picture \cite{2005JSMTE..04..010C,2007JSMTE..10....4C} as one more feature of holographic QFTs.
This picture was proposed to describe the time dependence of the entanglement entropy in two-dimensional conformal field theories ($2$d CFTs).
The entanglement entropy characterizes the bipartite entanglement between the subsystem under consideration and its complement.
This picture can describe the time dependence of the entanglement entropy in $2$d free field theories \cite{2016JHEP...11..166C,2018JSMTE..11.3103W,2015PhRvB..92g5109W,Castro-Alvaredo:2018dja,Parez:2020vsp,Capizzi:2022igy}.
Later, \cite{Asplund:2013zba,Asplund:2014coa} found that in the late-time regime of 2d holographic CFTs, the quasiparticle picture does not describe the time evolution of the entanglement entropy for the subsystem given by the multi-intervals.
The breakdown of the quasiparticle picture results in the quantum thermalization \cite{2022JHEP...06..100G}.
Some effective descriptions of the entanglement dynamics in the holographic CFTs are proposed \cite{2014PhRvL.112a1601L,2014PhRvD..89f6012L,2017PhRvX...7c1016N,2018arXiv180300089J,2020JHEP...01..031K}.
Another dynamical phenomenon expected to be induced by the strong scrambling ability is the logarithmic growth with time of the entanglement entropy \cite{2014arXiv1405.5946C,Asplund:2014coa,2018JHEP...01..115K,hart2021logarithmic}.
We start from the vacuum state, insert the local operator into the infinite interval system, and evolve it with a uniform Hamiltonian.
We choose the half infinite line as the subsystem.
In the quasiparticle picture for this system, the quasiparticles are induced by the local operator, they are entangled with each other, propagate to the left and right with the speed of light, and the entanglement entropy saturates at the value determined by the entanglement between them for the late times \cite{Nozaki:2014hna,Nozaki:2014uaa}.
This picture can describe the time-dependence of the entanglement entropy in $2$d non-holographic CFTs qualitatively \cite{PhysRevD.90.041701,2015PhRvD..92f5010C,2015PhRvD..92f5010C,2015JHEP...04..099G,2015JHEP...10..173C,2018JHEP...05..154G,2017JPhA...50e5002C} and its late-time value quantitatively \cite{Nozaki:2014hna,Nozaki:2014uaa,2014arXiv1405.5946C,2015JHEP...08..011C,2015JHEP...01..102C,2016PhRvD..93j5032C} even in higher dimensional systems than two \cite{Nozaki:2014hna,2016JHEP...02..150N,2016JHEP...12..069N,2017PhRvD..96b5019N}. 
In $2$d holographic CFTs, the quasiparticle picture can describe the early-time dependence of the entanglement entropy \cite{2014arXiv1405.5946C,Kusuki:2019gjs}. 
However, in the time region where the entanglement between quasiparticles contributes to the entanglement entropy, this picture does not describe the logarithmic increase with time in the entanglement entropy.
Therefore, this logarithmic growth may capture the unique feature of the dynamics induced by the holographic CFTs.
To investigate the details of this logarithmic growth in time, we considered the process induced by the local operator evolved by the time evolution constructed of the Euclidean and Lorentzian ones \cite{Mao:2024cnm}. 
Here, the Hamiltonian determining the Euclidean time evolution differs from the one determining the Lorentzian one.
In this case, there are two time orderings of Euclidean and Lorentzian time evolution: In the first case, we evolve the local operator with the Euclidean time evolution, and then do it with the Lorentzian one; in the second case, we evolve it with the Lorentzian one, and then do it with the Euclidean one.
Then, we found that the increase in the time in the entanglement entropy depends on the time orderings of the Euclidean and Lorentzian time evolution.
In this paper, to investigate the effect of time orderings (the time ordering effect) on the entanglement entropy in more detail, we will consider the Lorentzian time evolution collecting the quasiparticles around two spatial points, and then investigate the time ordering effect on the quantum entanglement between them.
Additionally, we will consider the process of destroying the entangled pair, and then will investigate if the entanglement entropy decreases when the quasiparticle decays in the spatial region far from the subsystem under consideration.
We will also investigate the time ordering effect on the mutual information and decrease in the mutual information due to the decay of the quasiparticle.
Furthermore, we will explore the gravity dual of the systems considered as in the papers \cite{Banados:1998gg,Roberts:2012aq,Fitzpatrick:2015zha,Abajian:2023bqv,Tian:2024fmo}.
\subsection*{Summary}
In this paper, we inserted a primary operator into the spatially-periodic system in the vacuum state as the resource of quantum entanglement, and then time-evolved it with the $2$d CFT Hamiltonian on the curved background. 
Then, by using the quantities characterizing the bipartite entanglement and non-local correlation, so-called entanglement entropy and mutual information, we explored the entanglement dynamics induced by this local operator. 
The evolution under consideration constitutes the Euclidean time evolution, the process of taming the divergence of the norm, and the Lorentzian time evolution, the process of evolving the systems with time.
The background of the Hamiltonian, inducing the Lorentzian time evolution, has two points where massless particles propagate. 
By changing the order of these processes, we explored the effect of the composite time evolution on the entanglement entropy and mutual information, and found that the significant difference is induced by the time ordering effect: 
When we evolve the system with Euclidean time evolution, and then evolve it with Lorentzian one, for the large Lorentzian time region, the entanglement entropy for the subsystem, containing the only single point where the massless particles propagate, logarithmically grows with the Lorentzian time forever.
When we change the time ordering of the Euclidean and Lorentzian time evolution, the entanglement entropy for the same subsystem saturates to a certain constant.

Furthermore, we investigated the relation between the energy density and quantum properties such as bipartite entanglement and non-local correlation in $2$d holographic CFT.
Then, we found the formula that explicitly presents the relation between them (for details, see Section \ref{sec:Relation-between-EE-and-energy}.).
By exploiting this formula, we investigated how the difference, induced by the time ordering effect, of the entanglement dynamics occurs. When the system is evolved with the Euclidean time evolution, and then is evolved with the Lorentzian one, the insertion of the local operator induces the excitation entirely distributed in the system. This well-distributed excitation induces the logarithmic growth of the entanglement entropy.
By contrast, when the system is evolved with the Lorentzian time evolution, and then is evolved with the Euclidean one, the local operator induces the local excitation, so that the entanglement entropy eventually saturates to a certain value.

We also investigated if the increase in the entanglement entropy is due to the entanglement between the quasiparticles.
To this end, we considered the process of destroying the quasiparticle contributing to the entanglement entropy, but not being contained in the subsystem where we calculated the entanglement entropy.
Then, we found that the entanglement entropy decays according to the decay of the quasiparticle in the region far from the subsystem under consideration. Furthermore, we found that the decay of the quasiparticles can induce the decay of the mutual information.

We also found the gravity dual of the systems considered in this paper. We consider the primary operator that possesses the large conformal dimension, where the gravitational dual of the system may be approximated by a black hole. We explored the dynamical behavior of the black hole's event horizon that corresponds to the primary operator with the large conformal dimension.  We found that according to the motion of quasiparticles, the shape of the black hole's event horizon changes. 
\subsection*{Organization of this paper}
Here, we will explain the organization of this paper.
In Section \ref{sec:setup}, we will explain the systems considered in this paper, describe the definition of the entanglement entropy and the mutual information, and then describe how to calculate the operator evolution induced by the Hamiltonian considered in this paper.
In Section \ref{sec:holographic-entanglement-entropy}, we will explain how to calculate the entanglement entropy in $2$d holographic CFTs, and present our findings.
In Section \ref{sec:mutual-information}, we will report on the time dependence of the mutual information for the systems under consideration.
In Section \ref{sec:gravity-dual}, we will present the gravity dual of the systems considered in this paper.
In Section \ref{sec:discussion-and-future-directions}, we will discuss our findings and comment on a few of the future problems.
\section{The setup \label{sec:setup}}
In this paper, we insert the local operator, as the source of entanglement, into the vacuum state of the two-dimensional conformal field theories ($2$d CFTs).
The systems with the insertion of local operators, under consideration, are in
\be
\ket{\Psi_{i}(x,t)}=\mathcal{N}_i\mathcal{O}_{H,i}(x,t)\ket{0},
\ee
where $i$ labels the states, $\mathcal{O}_{H,i}(x,t)$ is a local operator in the Heisenberg picture, normalization constant, $\mathcal{N}_i$, guarantees $\left \langle \Psi_i|\Psi_i \right \rangle=1$, and $\ket{0}$ is the vacuum of the $2$d CFT Hamiltonian on the flat space, 
\be
H_0=\int^L_0 dx e(x), 
\ee
where we assume that the system is spatially periodic with the period of $L$, and $e(x)$ is the energy density at $x$.
The Heisenberg operators under consideration are 
\be
\begin{split}
\mathcal{O}_{H,1}(x,t)&=e^{-iH_{\text{2-SSD}}t}e^{-\epsilon H_0}\mathcal{O}(x)e^{\epsilon H_0}e^{iH_{\text{2-SSD}}t},\\
\mathcal{O}_{H,2}(x,t)&=e^{-\epsilon H_0}e^{-iH_{\text{2-SSD}}t}\mathcal{O}(x)e^{iH_{\text{2-SSD}}t}e^{\epsilon H_0},\\
\mathcal{O}_{H,3}(x,t)&=e^{-\epsilon_2 H_{SSD}}e^{-iH_{\text{2-SSD}}t}e^{-\epsilon_1 H_0}\mathcal{O}(x)e^{\epsilon_1 H_0}e^{iH_{\text{2-SSD}}t}e^{\epsilon_2 H_{SSD}},\\
\mathcal{O}_{H,4}(x,t)&=e^{-\epsilon_2 H_{SSD}}e^{-\epsilon_1 H_0}e^{-iH_{\text{2-SSD}}t}\mathcal{O}(x)e^{iH_{\text{2-SSD}}t}e^{\epsilon_1 H_0}e^{\epsilon_2 H_{SSD}},\\
\end{split}
\ee
where $H_{\text{SSD}}$ and $H_{\text{2-SSD}}$ are CFT Hamiltonians on the curved background \cite{2018arXiv180500031W,PhysRevResearch.3.023044,PhysRevB.102.205125,PhysRevB.97.184309,2020PhRvX..10c1036F,2022arXiv221100040W}, 
\be
\begin{split}
   H_{\text{SSD}}=2\int^{L}_0dx \sin^2{\left(\f{ \pi x}{L}\right)} e(x),~H_{\text{2-SSD}}=2\int^{L}_0dx \sin^2{\left(\f{2 \pi x}{L}\right)} e(x).
\end{split}
\ee
Here, we assume that $\mathcal{O}$ is the primary operator with chiral and anti-chiral dimensions of $(h_\mathcal{O},\overline{h}_\mathcal{O})$, and 
the normalization constant, $\mathcal{N}_i$, is given by 
\be
\mathcal{N}_i=\left(\f{1}{\bra{0}\mathcal{O}^{\dagger}_{H,i}(x,t) \mathcal{O}_{H,i}(x,t)\ket{0}}\right)^{\f{1}{2}}.
\ee
The $2$d CFT Hamiltonian, $H_{\text{SSD}}$, is called sine-square-deformed Hamiltonian (SSD Hamiltonian).
This sine-square-deformation was originally introduced to remove the boundary effect in the finite spin system \cite{Gendiar01102009,Gendiar01022010,Hikihara_2011,2011PhRvA..83e2118G,Shibata:2011jup,Shibata_2011,Maruyama_2011,Katsura:2011zyx,Katsura:2011ss,PhysRevB.86.041108,PhysRevB.87.115128}, and then was generalized to $2$d CFTs \cite{Tada:2014kza,Ishibashi:2015jba,Ishibashi:2016bey, Okunishi_2016,2016PhRvB..93w5119W}. 
For further development about Floquet driving non-equilibrium phenomena and so on in terms of the SSD Hamiltonian, refer to \cite{PhysRevB.97.184309,2019JPhA...52X5401M,2021arXiv211214388G,PhysRevLett.118.260602,2018arXiv180500031W,2020PhRvX..10c1036F,Han_2020,2021PhRvR...3b3044W,2020arXiv201109491F,2021arXiv210910923W,PhysRevB.103.224303,PhysRevResearch.2.023085,Moosavi2021,PhysRevLett.122.020201,10.21468/SciPostPhys.3.3.019,2016arXiv161104591Z,2020PhRvR...2c3347R,HM,2018PhRvL.120u0604A,2019PhRvB..99j4308M,2022arXiv221100040W,2023arXiv230208009G,2023arXiv231019376N,2023arXiv230501019G,Das:2023xaw,2023arXiv230904665K,Liu:2023tiq,2024arXiv240315851M,2024arXiv240501642L,2024arXiv240216555B,2024arXiv240407884J,2024arXiv240806594B,2024JHEP...08..190M,2025arXiv250104795F,Das:2025wjo}.

We define the density operator for the systems under consideration as
\be
\rho_i(x,t)=\ket{\Psi_i(x,t)} \bra{\Psi_i(x,t)}.
\ee
Then, spatially divide the system into $A$ and $\overline{A}$, the space complement to $A$.
Define the reduced density matrix associated with $A$ as
\be
\rho_{i,A}(x,t)=\Tr_{\overline{A}} \rho_i(x,t).
\ee
To explore bipartite entanglement between $A$ and $\overline{A}$, define the $n$-th moment of R\'enyi entanglement entropy as
\be
\begin{split}
    S^{(n)}_A=\f{1}{1-n}\log{\left[\Tr_A\left(\rho^n_{i,A}(x,t)\right)\right]},
\end{split}
\ee
where $n$ is a positive integer.
Furthermore, define entanglement entropy as the Von Neumann entropy for $\rho_A$,
\be
S_A=\lim_{n\rightarrow 1} S^{(n)}_{A}=-\Tr_A \left(\rho_{i,A}(x,t)\log{\rho_{i,A}}(x,t)\right).
\ee
To study the increase in the bipartite entanglement due to the insertion of the local operator, define the increase in the entanglement entropy as 
\be \label {eq:increase-in-EE}
\Delta S_{A}(t)=S_{A}(t)-S_{A}(t=0).
\ee
Although the entanglement entropy should depend on the cutoff scale in the quantum field theories, $\Delta S_{A}(t)$ is independent of it.

Subsequently, spatially divide the system into $A$, $B$, and $\overline{A\cup B}$, the complement space to them, and then define the mutual information between $A$ and $B$ as 
\be \label{eq:def-of-MI}
I_{A,B}(t)=S_{A}(t)+S_{B}(t)-S_{A\cup B}(t),
\ee
where $S_{\alpha=A,B, A\cup B}$ denote the entanglement entropies associated with $A$, $B$, and $A\cup B$.
This quantity is cutoff-free, and characterizes the non-local correlation between $A$ and $B$.
\subsection{Twist operator formalism \label{sec:twist-op-formalism}}
To compute (\ref{eq:increase-in-EE}) and (\ref{eq:def-of-MI}), we start from the Euclidean density operator $\rho^{\text{Euc.}}_i(x,\tau)$, instead of $\rho_i(x,t)$.
Replacing $it$ of $\rho_{i}(x,t)$ with $\tau$, define $\rho^{\text{Euc.}}_i(x,\tau)$ as 
\be
\begin{split}
    \rho^{\text{Euc.}}_i(x,\tau) =\f{\mathcal{O}^{\text{Euc.}}_{H,i;+}(x,\tau)\ket{0}\bra{0}\tilde{\mathcal{O}}^{\text{Euc.}}_{H,i;-}(x,\tau)}{\bra{0}\tilde{\mathcal{O}}^{\text{Euc.}}_{H,i;-}(x,\tau)\mathcal{O}^{\text{Euc.}}_{H,i;+}(x,\tau)\ket{0}},
\end{split}
\ee
where $\mathcal{O}^{\text{Euc.}}_{H,i;\pm}(x,\tau)$ is defined as
\be \label{eq:Euclidena-Heisenberg}
\begin{split}
   &\mathcal{O}^{\text{Euc.}}_{H,1;+}(x,\tau)=\mathcal{O}^{\text{Euc.}}_{H,1}(x,\tau,\epsilon)=e^{-\tau H_{\text{2-SSD}}}e^{-\epsilon H_0}\mathcal{O}(x)e^{\epsilon H_0}e^{\tau H_{\text{2-SSD}}},\\
&\mathcal{O}^{\text{Euc.}}_{H,2;+}(x,\tau)=\mathcal{O}^{\text{Euc.}}_{H,2}(x,\tau,\epsilon)=e^{-\epsilon H_0}e^{-\tau H_{\text{2-SSD}}}\mathcal{O}(x)e^{\tau H_{\text{2-SSD}}}e^{\epsilon H_0},\\
&\mathcal{O}^{\text{Euc.}}_{H,3;+}(x,\tau)=\mathcal{O}^{\text{Euc.}}_{H,3}(x,\tau,\epsilon_1,\epsilon_2)=e^{-\epsilon_2 H_{SSD}}e^{-\tau H_{\text{2-SSD}}}e^{-\epsilon_1 H_0}\mathcal{O}(x)e^{\epsilon_1 H_0}e^{\tau H_{\text{2-SSD}}}e^{\epsilon_2 H_{SSD}},\\
& \mathcal{O}^{\text{Euc.}}_{H,4;+}(x,\tau)=\mathcal{O}^{\text{Euc.}}_{H,4}(x,\tau,\epsilon_1,\epsilon_2)=e^{-\epsilon_2 H_{SSD}}e^{-\epsilon_1 H_0}e^{-\tau H_{\text{2-SSD}}}\mathcal{O}(x)e^{\tau H_{\text{2-SSD}}}e^{\epsilon_1 H_0}e^{\epsilon_2 H_{SSD}},\\
&\mathcal{O}^{\text{Euc.}}_{H,i=1,2;-}(x,\tau)=\mathcal{O}^{\text{Euc.}}_{H,i=1,2}(x,\tau,-\epsilon),~\mathcal{O}^{\text{Euc.}}_{H,i=3,4;-}(x,\tau)=\mathcal{O}^{\text{Euc.}}_{H,4}(x,\tau,-\epsilon_1,-\epsilon_2).
\end{split}
\ee
The symbol, $\tilde{\mathcal{O}}$, means that we replace  $\mathcal{O}$ of (\ref{eq:Euclidena-Heisenberg}) with $\mathcal{O}^{\dagger}$.
Define a subsystem, $A$, as the spatial interval between $x=X_2$ and $x=X_1$, where $L>X_1>X_2>0$.
Then, define the Euclidean reduced density matrix associated with $A$ as $\rho^{\text{Euc.}}_{i,A}(x,\tau)=\Tr_{\overline{A}}\rho^{\text{Euc.}}_{i}(x,\tau)$, and  define the Euclidean R\'enyi entanglement entropy and von Neumann entropy as
\begin{equation}
S_{A, i}^{\text{Euc.}(n)}=\frac{1}{1-n} \log \left[\Tr_A\left( \rho^\text{Euc.}_{A, i}(x,\tau)\right)^n\right], \quad S_{A, i, E}=\lim _{n \rightarrow 1} S_{A, i}^{\text{Euc.}(n)}
\end{equation}
In the twist operator formalism \cite{Calabrese_2004,Calabrese_2009}, the Euclidean $n$-th R\'enyi entanglement entropy for $A$ is given by 
\begin{equation} \label{eq:Renyi-entanglement-entropy}
\begin{split}
   S^{\text{Euc.}(n)}_{A, i}=\frac{1}{1-n} \log \left[\frac{\left\langle \tilde{\mathcal{O}}^{\text{Euc.}}_{H,n,i;-}(x,\tau) \sigma_n\left(X_1\right) \bar{\sigma}_n\left(X_2\right) \mathcal{O}^{\text{Euc.}}_{H,n,i;+}(x,\tau) \right\rangle}{\left\langle  \tilde{\mathcal{O}}^{\text{Euc.}}_{H,n,i;-}(x,\tau)
  \mathcal{O}^{\text{Euc.}}_{H,n,i;+}(x,\tau) \right\rangle^n}\right],
\end{split}
\end{equation}
where $\langle\cdot\rangle$ denotes the vacuum expectation value, $\sigma_n$ and $\bar{\sigma}_n$ are the twist and anti-twist operators with the conformal dimensions, $\left(h_n, {\overline{h}}_n\right)=\left(\frac{c\left(n^2-1\right)}{24 n}, \frac{c\left(n^2-1\right)}{24 n}\right)$, and $\mathcal{O}^{\text{Euc.}}_{H,n,i;\pm}$ and $\tilde{\mathcal{O}}^{\text{Euc.}}_{H,n,i;\pm}$ are the primary operators with the conformal dimensions $\left(n h_{\mathcal{O}}, n h_{\mathcal{O}}\right)$ as in \cite{Asplund:2014coa}. 
By using (\ref{eq:Renyi-entanglement-entropy}), and then performing the analytic continuation, $\tau=it$, we will study the time dependence of $S_A$ and $S^{(n)}_A$ later.
\subsection{The Primary Operator in the Heisenberg picture \label{sec:PrimaryO.P.-in-Heisenberg-p}}
Here, we report on the details of the primary operators in the Heisenberg picture.  
We begin by considering $H_{\text{q-M\"obius}}$ which is defined as 
\be \label{eq:qMobius-Hamiltonian}
\begin{split}
   &H_{q-\text{M\"obius}} = \int^L_0 dx \left[1-\tanh{2\theta}\left(1-2\sin^2{\left(\f{q \pi x}{L}\right)}\right)\right]T(x)\\
    & = \int^L_0 \f{dw}{2 \pi i} \left[1-\tanh{2\theta}\left( \f{e^{\f{2 \pi q w}{L}}+e^{-\f{2 \pi q w}{L}}}{2}\right)\right]T(w)+\int^L_0 \f{d\overline{w}}{2 \pi i} \left[1-\tanh{2\theta}\left( \f{e^{\f{2 \pi q \overline{w}}{L}}+e^{-\f{2 \pi q \overline{w}}{L}}}{2}\right)\right]\overline{T}(\overline{w})\\
    & =\frac{2 q \pi}{ L \cosh (2 \theta)} \oint \f{\grave{z}}{2\pi i } T(\grave{z}) d\grave{z}+\frac{2 q \pi}{ L \cosh (2 \theta)} \oint \f{\overline{\grave{z}}}{2\pi i } T(\overline{\grave{z}}) d\overline{\grave{z}}-\frac{2\pi c}{12 L}\\
    &=\frac{2 q \pi}{L \cosh (2 \theta)} L_0^{(\grave{z})}+\frac{2 q \pi}{L \cosh (2 \theta)} \overline{L}_0^{(\overline{\grave{z}})}-\frac{2\pi c}{12 L}, 
\end{split}
\ee
where $q$ is a positive integer and the pairs of complex coordinates  are introduced as follows:
\be
\begin{split}
    &(\grave{z},\overline{\grave{z}})=(f(\tilde{z}),\overline{f}(\overline{\tilde{z}}))=\left(-\frac{\cosh \theta \tilde{z}-\sinh \theta}{\sinh \theta \tilde{z}-\cosh \theta}, -\frac{\cosh \theta \overline{\tilde{z}}-\sinh \theta}{\sinh \theta \overline{\tilde{z}}-\cosh \theta}\right), \\
&(\tilde{z},\bar{\tilde{z}})=\left(e^{\frac{2 q \pi w}{L}}, e^{\frac{2 q \pi \overline{w}}{L}}\right), 
(z, \bar{z})=\left(e^{\frac{2 \pi(i x+\tau)}{L}}, e^{\frac{2 \pi(-i x+\tau)}{L}}\right)=\left(e^{\frac{2 \pi w}{L}}, e^{\frac{2 \pi \overline{w}}{L}}\right).
\end{split}
\ee
In the last line of (\ref{eq:qMobius-Hamiltonian}),
we define the Virasoro generators in $(\grave{z},\overline{\grave{z}})$ as 
\be
L^{(\grave{z})}_{n}=\oint \f{d \grave{z}}{2\pi i}\grave{z}^{n+1}T(\grave{z}), \overline{L}^{(\overline{\grave{z}})}_{n}=\oint \f{d \overline{\grave{z}}}{2\pi i}\overline{\grave{z}}^{n+1}\overline{T}(\overline{\grave{z}}).
\ee

In the pair of the complex coordinates, $(\grave{z},\overline{\grave{z}})$, the transformation induced by $H_{q-\text{M\"obius}}$ acts on the primary operator with $(h_{\mathcal{T}},h_{\mathcal{T}})$, $\mathcal{T}$, as a dilatation operator, 
\begin{equation}
e^{H_{q-\text{M\"obius}} \tau} \mathcal{T}(\grave{z}, \overline{\grave{z}}) e^{-H_{q-\text{M\"obius}} \tau}=\lambda^{2 h_{\mathcal{T}}} \mathcal{T}(\lambda \grave{z}, \lambda \overline{\grave{z}}),
\end{equation}
where $\lambda$ is defined as 
\be \label{eq:scaling-factor}
\lambda:=\exp \left(\frac{2 \pi q \tau}{L \cosh (2 \theta)}\right) .
\ee
Let $(z^{\text{new}}_{\theta}, \overline{z}^{\text{new}}_{\theta})$ denote the new position of the primary operator $\mathcal{T}$ in the $(z,\overline{z})$ coordinates after the action of $H_{q\text{-M\"obius}}$. This position is determined by the following identification condition:
\be \label{eq:identification}
f\left(\left(z^{\text {new }}_{\theta}\right)^q\right)=\lambda f(z^q), \quad \overline{f}\left(\left(\overline{z}^{\text {new }}_{\theta}\right)^q\right)=\lambda \overline{f}(\overline{z}^q).
\ee
Using this, the coordinates $(z^{\text{new}}_{\theta}, \overline{z}^{\text{new}}_{\theta})$ are related to the $(w, \overline{w})$ coordinates via
\be
(z^{\text{new}}_{\theta}, \overline{z}^{\text{new}}_{\theta}) = \left(e^{\frac{2\pi w^{\text{new}}_{\theta}}{L}}, e^{\frac{2\pi \overline{w}^{\text{new}}_{\theta}}{L}}\right),
\ee
where the explicit expressions are given by
\begin{equation}
\begin{split}
    z^{\text{new}}_{\theta} &= \left(\frac{[(1-\lambda) \cosh 2\theta - (\lambda+1)] z^q + (\lambda-1) \sinh 2\theta}{(1-\lambda) \sinh 2\theta \, z^q + [(\lambda - 1) \cosh 2\theta - (\lambda + 1)]}\right)^{1/q}, \\
    \overline{z}^{\text{new}}_{\theta} &= \left(\frac{[(1-\lambda) \cosh 2\theta - (\lambda+1)] \overline{z}^q + (\lambda-1) \sinh 2\theta}{(1-\lambda) \sinh 2\theta \, \overline{z}^q + [(\lambda - 1) \cosh 2\theta - (\lambda + 1)]}\right)^{1/q}, \\
    w^{\text{new}}_{\theta} &= \frac{L}{2\pi q} \log \left[\frac{[(1-\lambda) \cosh 2\theta - (\lambda+1)] z^q + (\lambda - 1) \sinh 2\theta}{(1-\lambda) \sinh 2\theta \, z^q + [(\lambda - 1) \cosh 2\theta - (\lambda + 1)]}\right], \\
    \overline{w}^{\text{new}}_{\theta} &= \frac{L}{2\pi q} \log \left[\frac{[(1-\lambda) \cosh 2\theta - (\lambda+1)] \overline{z}^q + (\lambda - 1) \sinh 2\theta}{(1-\lambda) \sinh 2\theta \, \overline{z}^q + [(\lambda - 1) \cosh 2\theta - (\lambda + 1)]}\right].
\end{split}
\end{equation}
Define 
$(z^{\text{new}},\overline{z}^{\text{new}})=(e^{\f{2\pi w^{\text{new}}}{L}},e^{\f{2\pi \overline{w}^{\text{new}}}{L}})$ as the leading term of $(z^{\text{new}}_{\theta}, \overline{z}^{\text{new}}_{\theta})$ in the large $\theta$ expansion, $\theta \gg 1$,
\be
\begin{split}
    &z^{\text {new }}=\left(\frac{-\pi  q \tau +(\pi  q \tau -L) z^q}{\pi  q \tau  z^q-L-\pi  q \tau }\right)^{1/q},
    \overline{z}^{\text {new }}=\left(\frac{-\pi  q \tau +(\pi  q \tau -L) \overline{z}^q}{\pi  q \tau  \overline{z}^q-L-\pi  q \tau }\right)^{1/q},\\
    &w^{\text{new}}= \f{L}{2\pi q} \log{\left(\frac{-\pi  q \tau +(\pi  q \tau -L) z^q}{\pi  q \tau  z^q-L-\pi  q \tau }\right)},~\overline{w}^{\text{new}}=\f{L}{2\pi q} \log{\left(\frac{-\pi  q \tau +(\pi  q \tau -L) \overline{z}^q}{\pi  q \tau  \overline{z}^q-L-\pi  q \tau }\right)}
\end{split}
\ee
Define Virasoro generators in $(z,\overline{z})$ as 
\be
L^{z}_{n}=\oint \f{d z}{2\pi i}z^{n+1}T(z), \overline{L}^{\overline{z}}_{n}=\oint \f{d \overline{z}}{2\pi i}\overline{z}^{n+1}\overline{T}(\overline{z}).
\ee
The Hamiltonians under consideration are given by
\be
\begin{aligned}
& H_0=\frac{2 \pi}{L}\left(L_0+\bar{L}_0\right) -\frac{\pi c}{6 L},\\
& H_{\text {SSD }}=\frac{2 \pi}{L}\left[L_0+\bar{L}_0-\frac{1}{2}\left(L_1+L_{-1}+\bar{L}_1+\bar{L}_{-1}\right)\right]-\frac{\pi c}{6 L},\\
& H_{\text {2-SSD }}=\frac{2 \pi}{L}\left[L_0+\bar{L}_0-\frac{1}{2}\left(L_2+L_{-2}+\bar{L}_2+\bar{L}_{-2}\right)\right]-\frac{\pi c}{6 L}.\\
\end{aligned}
\ee
Then, the primary operator considered in (\ref{eq:Euclidena-Heisenberg}) results in the following transformation
\be\label{destroy-set-up}
\begin{split}
\mathcal{O}^{\text{Euc.}}_{H,1;+}(x,\tau)&=e^{-iH_{\text{2-SSD}}t}e^{-\epsilon H_0}\mathcal{O}\left(w_x, \overline{w}_x\right)e^{\epsilon H_0}e^{iH_{\text{2-SSD}}t}  =\left(\frac{d w_{-\epsilon}^{\text {new},1}}{d w_x}\right)^{h_\mathcal{O}}\left(\frac{d \overline{w}_{-\epsilon}^{\text {new},1}}{d \overline{w}_x}\right)^{\overline{h}_\mathcal{O}} \mathcal{O}\left(w_{-\epsilon}^{\text {new},1}, \overline{w}_{-\epsilon}^{\text {new},1}\right),\\
\mathcal{O}^{\text{Euc.}}_{H,2;+}(x,\tau)&=e^{-\epsilon H_0} e^{-iH_{\text{2-SSD}}t} \mathcal{O}\left(w_x, \overline{w}_x\right) e^{iH_{\text{2-SSD}}t} e^{\epsilon H_0}=\left(\frac{d w_{-\epsilon}^{\text {new},2}}{d w_x}\right)^{h_\mathcal{O}}\left(\frac{d \overline{w}_{-\epsilon}^{\text {new},2}}{d \overline{w}_x}\right)^{\overline{h}_\mathcal{O}} \mathcal{O}\left(w_{-\epsilon}^{\text {new},2}, \overline{w}_{-\epsilon}^{\text {new},2}\right),\\
\mathcal{O}^{\text{Euc.}}_{H,3;+}(x,\tau)&=e^{-\epsilon_2 H_{\text{SSD}}}e^{-iH_{\text{2-SSD}}t}e^{-\epsilon_1 H_0}\mathcal{O}\left(w_x, \overline{w}_x\right)e^{\epsilon_1 H_0}e^{iH_{\text{2-SSD}}t} e^{\epsilon_2 H_{\text{SSD}}} \\ &=\left(\frac{d w_{-\epsilon}^{\text {new},3}}{d w_x}\right)^{h_\mathcal{O}}\left(\frac{d \overline{w}_{-\epsilon}^{\text {new},3}}{d \overline{w}_x}\right)^{\overline{h}_\mathcal{O}} \mathcal{O}\left(w_{-\epsilon}^{\text {new},3}, \overline{w}_{-\epsilon}^{\text {new},3}\right),\\
\mathcal{O}^{\text{Euc.}}_{H,4;+}(x,\tau)&=e^{-\epsilon_2 H_{\text{SSD}}}e^{-\epsilon_1 H_0}e^{-iH_{\text{2-SSD}}t}\mathcal{O}\left(w_x, \overline{w}_x\right)e^{iH_{\text{2-SSD}}t} e^{\epsilon_1 H_0} e^{\epsilon_2 H_{\text{SSD}}} \\ &=\left(\frac{d w_{-\epsilon}^{\text {new},4}}{d w_x}\right)^{h_\mathcal{O}}\left(\frac{d \overline{w}_{-\epsilon}^{\text {new},4}}{d \overline{w}_x}\right)^{\overline{h}_\mathcal{O}} \mathcal{O}\left(w_{-\epsilon}^{\text {new},4}, \overline{w}_{-\epsilon}^{\text {new},4}\right),\\
\tilde{\mathcal{O}}^{\text{Euc.}}_{H,1;-}(x,\tau)&=e^{-iH_{\text{2-SSD}}t}e^{\epsilon H_0}\mathcal{O}^{\dagger}\left(w_x, \overline{w}_x\right)e^{-\epsilon H_0}e^{iH_{\text{2-SSD}}t}  =\left(\frac{d w_\epsilon^{\text {new},1}}{d w_x}\right)^{h_\mathcal{O}}\left(\frac{d \overline{w}_\epsilon^{\text {new},1}}{d \overline{w}_x}\right)^{\overline{h}_\mathcal{O}} \mathcal{O}^{\dagger}\left(w_\epsilon^{\text {new},1}, \overline{w}_\epsilon^{\text {new},1}\right),\\
\tilde{\mathcal{O}}^{\text{Euc.}}_{H,2;-}(x,\tau)&=e^{\epsilon H_0} e^{-iH_{\text{2-SSD}}t} \mathcal{O}^{\dagger}\left(w_x, \overline{w}_x\right) e^{iH_{\text{2-SSD}}t} e^{-\epsilon H_0}=\left(\frac{d w_\epsilon^{\text {new},2}}{d w_x}\right)^{h_\mathcal{O}}\left(\frac{d \overline{w}_\epsilon^{\text {new},2}}{d \overline{w}_x}\right)^{\overline{h}_\mathcal{O}} \mathcal{O}^{\dagger}\left(w_\epsilon^{\text {new},2}, \overline{w}_\epsilon^{\text {new},2}\right),\\
\tilde{\mathcal{O}}^{\text{Euc.}}_{H,3;-}(x,\tau)&=e^{\epsilon_2 H_{\text{SSD}}}e^{-iH_{\text{2-SSD}}t}e^{\epsilon_1 H_0}\mathcal{O}\left(w_x, \overline{w}_x\right)e^{-\epsilon_1 H_0}e^{iH_{\text{2-SSD}}t} e^{-\epsilon_2 H_{\text{SSD}}} \\ &=\left(\frac{d w_{\epsilon}^{\text {new},3}}{d w_x}\right)^{h_\mathcal{O}}\left(\frac{d \overline{w}_{\epsilon}^{\text {new},3}}{d \overline{w}_x}\right)^{\overline{h}_\mathcal{O}} \mathcal{O}\left(w_{\epsilon}^{\text {new},3}, \overline{w}_{\epsilon}^{\text {new},3}\right),\\\\
\tilde{\mathcal{O}}^{\text{Euc.}}_{H,4;-}(x,\tau)&=e^{\epsilon_2 H_{\text{SSD}}}e^{\epsilon_1 H_0}e^{-iH_{\text{2-SSD}}t}\mathcal{O}\left(w_x, \overline{w}_x\right)e^{iH_{\text{2-SSD}}t} e^{-\epsilon_1 H_0} e^{-\epsilon_2 H_{\text{SSD}}} \\ &=\left(\frac{d w_{\epsilon}^{\text {new},4}}{d w_x}\right)^{h_\mathcal{O}}\left(\frac{d \overline{w}_{\epsilon}^{\text {new},4}}{d \overline{w}_x}\right)^{\overline{h}_\mathcal{O}} \mathcal{O}\left(w_{\epsilon}^{\text {new},4}, \overline{w}_{\epsilon}^{\text {new},4}\right).\\
\end{split}
\ee
where $w_x=i x, \overline{w}_x=-i x$, and $w_{\pm \epsilon}^{\text {new},i}$ and $\overline{w}_{\pm \epsilon}^{\text {new},i}$ are given by
\be
\begin{split}
&w_{-\epsilon}^{\text {new},1}=\frac{L}{4 \pi} \log \left[\frac{2 \pi  \tau  e^{\frac{4 \pi  \epsilon }{L}}+(L-2 \pi  \tau ) e^{\frac{4 i \pi  x}{L}}}{(L+2 \pi  \tau ) e^{\frac{4 \pi  \epsilon }{L}}-2 \pi  \tau  e^{\frac{4 i \pi  x}{L}}}\right],
w_{-\epsilon}^{\text {new},2}=-\epsilon+\frac{L}{4 \pi} \log \left[\frac{2 \pi  \tau +(L-2 \pi  \tau ) e^{\frac{4 i \pi  x}{L}}}{L-2 \pi  \tau  \left(-1+e^{\frac{4 i \pi  x}{L}}\right)}\right],\\
&w_{\epsilon}^{\text {new},1}=\frac{L}{4 \pi} \log \left[\frac{2 \pi  \tau  e^{\frac{-4 \pi  \epsilon }{L}}+(L-2 \pi  \tau ) e^{\frac{4 i \pi  x}{L}}}{(L+2 \pi  \tau ) e^{\frac{-4 \pi  \epsilon }{L}}-2 \pi  \tau  e^{\frac{4 i \pi  x}{L}}}\right],
w_{\epsilon}^{\text {new},2}=\epsilon+\frac{L}{4 \pi} \log \left[\frac{2 \pi  \tau +(L-2 \pi  \tau ) e^{\frac{4 i \pi  x}{L}}}{L-2 \pi  \tau  \left(-1+e^{\frac{4 i \pi  x}{L}}\right)}\right],\\
&\overline{w}_{-\epsilon}^{\text {new},1}=\frac{L}{4 \pi} \log \left[\frac{2 \pi  \tau  e^{\frac{4 \pi  \epsilon }{L}}+(L-2 \pi  \tau ) e^{\frac{-4 i \pi  x}{L}}}{(L+2 \pi  \tau ) e^{\frac{4 \pi  \epsilon }{L}}-2 \pi  \tau  e^{\frac{-4 i \pi  x}{L}}}\right],
\overline{w}_{-\epsilon}^{\text {new},2}=-\epsilon+\frac{L}{4 \pi} \log \left[\frac{2 \pi  \tau +(L-2 \pi  \tau ) e^{\frac{-4 i \pi  x}{L}}}{L-2 \pi  \tau  \left(-1+e^{\frac{-4 i \pi  x}{L}}\right)}\right],\\
&\overline{w}_{\epsilon}^{\text {new},1}=\frac{L}{4 \pi} \log \left[\frac{2 \pi  \tau  e^{\frac{-4 \pi  \epsilon }{L}}+(L-2 \pi  \tau ) e^{\frac{-4 i \pi  x}{L}}}{(L+2 \pi  \tau ) e^{\frac{-4 \pi  \epsilon }{L}}-2 \pi  \tau  e^{\frac{-4 i \pi  x}{L}}}\right],
\overline{w}_{\epsilon}^{\text {new},2}=\epsilon+\frac{L}{4 \pi} \log \left[\frac{2 \pi  \tau +(L-2 \pi  \tau ) e^{\frac{-4 i \pi  x}{L}}}{L-2 \pi  \tau  \left(-1+e^{\frac{-4 i \pi  x}{L}}\right)}\right],\\
\end{split}
\ee
\be
\begin{split}
&w_{-\epsilon}^{\text {new},3}=\frac{L}{2 \pi } \log \left[-\frac{-\pi  \epsilon _2+\left(\pi  \epsilon _2-L\right) \sqrt{\frac{2 \pi  \tau  e^{\frac{4 \pi  \epsilon _1}{L}}+(L-2 \pi  \tau ) e^{\frac{4 i \pi  x}{L}}}{(L+2 \pi  \tau ) e^{\frac{4 \pi  \epsilon _1}{L}}-2 \pi  \tau  e^{\frac{4 i \pi  x}{L}}}}}{L+\epsilon _2 \left(\pi -\pi  \sqrt{\frac{2 \pi  \tau  e^{\frac{4 \pi  \epsilon _1}{L}}+(L-2 \pi  \tau ) e^{\frac{4 i \pi  x}{L}}}{(L+2 \pi  \tau ) e^{\frac{4 \pi  \epsilon _1}{L}}-2 \pi  \tau  e^{\frac{4 i \pi  x}{L}}}}\right)}\right],\\
&w_{-\epsilon}^{\text {new},4}=\frac{L}{2 \pi } \log \left[-\frac{-\pi  \epsilon _2+e^{-\frac{2 \pi  \epsilon _1}{L}} \left(\pi  \epsilon _2-L\right) \sqrt{\frac{2 \pi  \tau +(L-2 \pi  \tau ) e^{\frac{4 i \pi  x}{L}}}{L-2 \pi  \tau  \left(-1+e^{\frac{4 i \pi  x}{L}}\right)}}}{-\pi  \epsilon _2 e^{-\frac{2 \pi  \epsilon _1}{L}} \sqrt{\frac{2 \pi  \tau +(L-2 \pi  \tau ) e^{\frac{4 i \pi  x}{L}}}{L-2 \pi  \tau  \left(-1+e^{\frac{4 i \pi  x}{L}}\right)}}+L+\pi  \epsilon _2}\right],\\
&w_{\epsilon}^{\text {new},3}=\frac{L}{2 \pi } \log \left[-\frac{\pi  \epsilon _2+\left(-L-\pi  \epsilon _2\right) \sqrt{\frac{2 \pi  \tau  e^{-\frac{4 \pi  \epsilon _1}{L}}+(L-2 \pi  \tau ) e^{\frac{4 i \pi  x}{L}}}{(L+2 \pi  \tau ) e^{-\frac{4 \pi  \epsilon _1}{L}}-2 \pi  \tau  e^{\frac{4 i \pi  x}{L}}}}}{L-\epsilon _2 \left(\pi -\pi  \sqrt{\frac{2 \pi  \tau  e^{-\frac{4 \pi  \epsilon _1}{L}}+(L-2 \pi  \tau ) e^{\frac{4 i \pi  x}{L}}}{(L+2 \pi  \tau ) e^{-\frac{4 \pi  \epsilon _1}{L}}-2 \pi  \tau  e^{\frac{4 i \pi  x}{L}}}}\right)}\right],\\
&w_{\epsilon}^{\text {new},4}=\frac{L}{2 \pi } \log \left[-\frac{-\pi  \epsilon _2+e^{-\frac{2 \pi  \epsilon _1}{L}} \left(\pi  \epsilon _2-L\right) \sqrt{\frac{2 \pi  \tau +(L-2 \pi  \tau ) e^{\frac{4 i \pi  x}{L}}}{L-2 \pi  \tau  \left(-1+e^{\frac{4 i \pi  x}{L}}\right)}}}{-\pi  \epsilon _2 e^{-\frac{2 \pi  \epsilon _1}{L}} \sqrt{\frac{2 \pi  \tau +(L-2 \pi  \tau ) e^{\frac{4 i \pi  x}{L}}}{L-2 \pi  \tau  \left(-1+e^{\frac{4 i \pi  x}{L}}\right)}}+L+\pi  \epsilon _2}\right],\\
\end{split}
\ee
\be
\begin{split}
&\overline{w}_{-\epsilon}^{\text {new},3}=\frac{L}{2 \pi } \log \left[-\frac{-\pi  \epsilon _2-\left(\pi  \epsilon _2-L\right) \sqrt{\frac{2 \pi  \tau  e^{\frac{4 \pi  \epsilon _1}{L}}+(L-2 \pi  \tau ) e^{-\frac{4 i \pi  x}{L}}}{(L+2 \pi  \tau ) e^{\frac{4 \pi  \epsilon _1}{L}}-2 \pi  \tau  e^{-\frac{4 i \pi  x}{L}}}}}{L+\epsilon _2 \left(\pi +\pi  \sqrt{\frac{2 \pi  \tau  e^{\frac{4 \pi  \epsilon _1}{L}}+(L-2 \pi  \tau ) e^{-\frac{4 i \pi  x}{L}}}{(L+2 \pi  \tau ) e^{\frac{4 \pi  \epsilon _1}{L}}-2 \pi  \tau  e^{-\frac{4 i \pi  x}{L}}}}\right)}\right],\\
&\overline{w}_{-\epsilon}^{\text {new},4}=\frac{L}{2 \pi } \log \left[-\frac{-\pi  \epsilon _2-e^{-\frac{2 \pi  \epsilon _1}{L}} \left(\pi  \epsilon _2-L\right) \sqrt{\frac{2 \pi  \tau +(L-2 \pi  \tau ) e^{-\frac{4 i \pi  x}{L}}}{L-2 \pi  \tau  \left(-1+e^{-\frac{4 i \pi  x}{L}}\right)}}}{\pi  \epsilon _2 e^{-\frac{2 \pi  \epsilon _1}{L}} \sqrt{\frac{2 \pi  \tau +(L-2 \pi  \tau ) e^{-\frac{4 i \pi  x}{L}}}{L-2 \pi  \tau  \left(-1+e^{-\frac{4 i \pi  x}{L}}\right)}}+L+\pi  \epsilon _2}\right],\\
&\overline{w}_{\epsilon}^{\text {new},3}=\frac{L}{2 \pi } \log \left[-\frac{\pi  \epsilon _2-\left(-L-\pi  \epsilon _2\right) \sqrt{\frac{2 \pi  \tau  e^{-\frac{4 \pi  \epsilon _1}{L}}+(L-2 \pi  \tau ) e^{-\frac{4 i \pi  x}{L}}}{(L+2 \pi  \tau ) e^{-\frac{4 \pi  \epsilon _1}{L}}-2 \pi  \tau  e^{-\frac{4 i \pi  x}{L}}}}}{L-\epsilon _2 \left(\pi +\pi  \sqrt{\frac{2 \pi  \tau  e^{-\frac{4 \pi  \epsilon _1}{L}}+(L-2 \pi  \tau ) e^{-\frac{4 i \pi  x}{L}}}{(L+2 \pi  \tau ) e^{-\frac{4 \pi  \epsilon _1}{L}}-2 \pi  \tau  e^{-\frac{4 i \pi  x}{L}}}}\right)}\right],\\
&\overline{w}_{\epsilon}^{\text {new},4}=\frac{L}{2 \pi } \log \left[-\frac{\pi  \epsilon _2-e^{\frac{2 \pi  \epsilon _1}{L}} \left(-L-\pi  \epsilon _2\right) \sqrt{\frac{2 \pi  \tau +(L-2 \pi  \tau ) e^{-\frac{4 i \pi  x}{L}}}{L-2 \pi  \tau  \left(-1+e^{-\frac{4 i \pi  x}{L}}\right)}}}{-\pi  \epsilon _2 e^{\frac{2 \pi  \epsilon _1}{L}} \sqrt{\frac{2 \pi  \tau +(L-2 \pi  \tau ) e^{-\frac{4 i \pi  x}{L}}}{L-2 \pi  \tau  \left(-1+e^{-\frac{4 i \pi  x}{L}}\right)}}+L-\pi  \epsilon _2}\right].\\
\end{split}
\ee

\section{Holographic Entanglement Entropy \label{sec:holographic-entanglement-entropy}}

In this section, we explore the time dependence of the entanglement entropy in $2$d holographic CFTs.
We begin with the $n$-th moment of the Euclidean R\'enyi entanglement entropy in (\ref{eq:Renyi-entanglement-entropy}).
After the transformation discussed in Section \ref{sec:PrimaryO.P.-in-Heisenberg-p}, the $n$-th moment of R\'enyi entanglement entropy results in
\be \label{eq:entanglement-entropy-fomula-1}
\begin{split}
 S^{(n)}_{A,i,E} =\f{1}{1-n}\log{\left[\f{\left \langle \tilde{\mathcal{O}}^{\text{Euc.}}_{H,n,i;-}(w_{\epsilon}^{\text {new}, i},\overline{w}_{\epsilon}^{\text {new}, i})\sigma_{n}(w_{X_1},\overline{w}_{X_1})\overline{\sigma}_n(w_{X_2},\overline{w}_{X_2})\mathcal{O}^{\text{Euc.}}_{H,n,i;+}(w_{-\epsilon}^{\text {new }, i},\overline{w}_{-\epsilon}^{\text {new}, i})\right \rangle}{\left \langle \tilde{\mathcal{O}}^{\text{Euc.}}_{H,n,i;-}(w_{\epsilon}^{\text {new}, i},\overline{w}_{\epsilon}^{\text {new}, i}) \mathcal{O}^{\text{Euc.}}_{H,n,i;+}(w_{-\epsilon}^{\text {new}, i},\overline{w}_{-\epsilon}^{\text {new}, i})\right \rangle^n}\right]}.
\end{split}
\ee
Thus, the contribution from the conformal factors is canceled out. By mapping from the cylinder to the complex plane, $(z, \bar{z})=\left(e^{\frac{2 \pi w}{L}}, e^{\frac{2 \pi \overline{w}}{L}}\right)$, we obtain $S_{A, i, E}^{(n)}$ as
\begin{equation}\label{eq:entanglement-entropy-fomula-2}
\begin{aligned}
S_{A, i, E}^{(n)} = & \f{1}{1-n}\log{\left[\f{\left \langle \tilde{\mathcal{O}}^{\text{Euc.}}_{H,n,i;-}(z_{\epsilon}^{\text {new}, i},\overline{z}_{\epsilon}^{\text {new}, i})\sigma_{n}(z_{X_1},\overline{z}_{X_1})\overline{\sigma}_n(z_{X_2},\overline{z}_{X_2})\mathcal{O}^{\text{Euc.}}_{H,n,i;+}(z_{-\epsilon}^{\text {new}, i},\overline{z}_{-\epsilon}^{\text {new}, i})\right \rangle}{\left \langle \tilde{\mathcal{O}}^{\text{Euc.}}_{H,n,i;-}(z_{\epsilon}^{\text {new}, i},\overline{z}_{\epsilon}^{\text {new}, i}) \mathcal{O}^{\text{Euc.}}_{H,n,i;+}(z_{-\epsilon}^{\text {new}, i},\overline{z}_{-\epsilon}^{\text {new}, i})\right \rangle^n}\right]}\\
& -\frac{c(1+n)}{24 n} \log \left[\prod_{i=1,2}\left|\frac{d z_{X_i}}{d w_{X_i}}\right|^2\right],
\end{aligned}
\end{equation}
where $z_{\pm \epsilon}^{\text {new}, i=1, 2}=e^{\frac{2 \pi w_{\pm \epsilon}^{\text {new}, i=1, 2}}{L}}, z_{X_{i=1,2}}=e^{\frac{2 \pi w_{X_{i=1,2}}}{L}}$ and $\bar{z}$ is the complex conjugate of $z$.
Subsequently, we perform the conformal map,
\begin{equation}\label{conformal-map-1}
\left(z^{\prime}, \overline{z}^{\prime}\right)=\left(\frac{\left(z_{\epsilon}^{\text {new }, i}-z\right)\left(z_{X_2}-z_{-\epsilon}^{\text {new },i}\right)}{\left(z-z_{-\epsilon}^{\text {new }, i}\right)\left(z_{\epsilon}^{\text {new }, i}-z_{X_2}\right)}, \frac{\left(\bar{z}_{\epsilon}^{\text {new }, i}-\bar{z}\right)\left(\bar{z}_{X_2}-\bar{z}_{-\epsilon}^{\text {new }, i}\right)}{\left(\bar{z}-\bar{z}_{-\epsilon}^{\text {new }, i}\right)\left(\bar{z}_{\epsilon}^{\text {new }, i}-\bar{z}_{X_2}\right)}\right).
\end{equation}
This maps the location of $\mathcal{O}_{n}$ and  $\tilde{\mathcal{O}}_{n}$ to $\infty$ and $0$, respectively, and then (\ref{eq:entanglement-entropy-fomula-2}) results in
\be
S_{A, i, E}^{(n)}=\frac{1}{1-n} \log \left[\left|z_{X_1}-z_{X_2}\right|^{-4 n h_n}\left|1-z_{c, i}\right|^{4 n h_n} G_n\left(z_{c, i}, \bar{z}_{c, i}\right)\right]-\frac{c(n+1)}{6 n} \log \left(\frac{2 \pi}{L}\right),
\ee
where the function of the cross ratios $\left(z_{c, i}, \bar{z}_{c, i}\right)$ is given by
\begin{equation}
\begin{aligned}
G_n\left(z_{c, i}, \bar{z}_{c, i}\right)
&=\left \langle \tilde{\mathcal{O}}^{\text{Euc.}}_{H,n,i;-}(0,0)\sigma_{n}\left(z_{c, i}, \bar{z}_{c, i}\right)\overline{\sigma}_n(1,1)\mathcal{O}^{\text{Euc.}}_{H,n,i;+}(\infty,\infty)\right \rangle,
\end{aligned}
\end{equation}
where 
the cross ratio is defined as 
\begin{equation}
\left(z_{c, i}, \bar{z}_{c, i}\right)=\left(\frac{\left(z_{\epsilon}^{\text {new }, i}-z_{X_1}\right)\left(z_{X_2}-z_{-\epsilon}^{\text {new },i}\right)}{\left(z_{X_1}-z_{-\epsilon}^{\text {new }, i}\right)\left(z_{\epsilon}^{\text {new }, i}-z_{X_2}\right)}, \frac{\left(\bar{z}_{\epsilon}^{\text {new }, i}-\bar{z}_{X_1}\right)\left(\bar{z}_{X_2}-\bar{z}_{-\epsilon}^{\text {new }, i}\right)}{\left(\bar{z}_{X_1}-\bar{z}_{-\epsilon}^{\text {new }, i}\right)\left(\bar{z}_{\epsilon}^{\text {new }, i}-\bar{z}_{X_2}\right)}\right).
\end{equation}
Subsequently, we closely look at the behavior of $\log G_n\left(z_{c, i}, \bar{z}_{c, i}\right)$ at the leading order of the small $n$ limit, where $n \approx 1$, with $z_{c, i}, \bar{z}_{c, i} \approx 1$. 
In this limit, the leading term of $\log G_n\left(z_{c, i}, \bar{z}_{c, i}\right)$ is given by
\be\label{eq:G_n}
\log G_n\left(z_{c, i}, \bar{z}_{c, i}\right) \approx \frac{c(1-n)}{6} \log \left[\frac{z_{c, i}^{\frac{\left(1-\alpha_{\mathcal{O}}\right)}{2}} \bar{z}_{c, i}^{\frac{\left(1-\bar{\alpha}_{\mathcal{O}}\right)}{2}}\left(1-z_{c, i}^{\alpha_{\mathcal{O}}}\right)\left(1-\bar{z}_{c, i}^{\bar{\alpha}_{\mathcal{O}}}\right)}{\alpha_{\mathcal{O}} \bar{\alpha}_{\mathcal{O}}}\right],
\ee
where $\alpha_{\mathcal{O}}=\bar{\alpha}_{\mathcal{O}}=\sqrt{1-24 h_{\mathcal{O}} / c}$. Consequently, in the Von Neumann limit, $n \rightarrow 1$, $S_{A, i, E}^{(n)}$ reduces to
\be \label{eq:entanglement-formula-in-Euclidean}
\begin{aligned}
\lim_{n\rightarrow 1} S^{(n)}_{A,i,E}=S_{A, i, E}= & \frac{c}{6} \log \left[\frac{z_{c, i}^{\frac{1-\alpha_{\mathcal{O}}}{{ }^2}} \bar{z}_{c, i}^{\frac{1-\bar{\alpha}_{\mathcal{O}}}{2}}\left(1-z_{c, i}^{\alpha_{\mathcal{O}}}\right)\left(1-\bar{z}_{c, i}^{\bar{\alpha}_{\mathcal{O}}}\right)}{\alpha_{\mathcal{O}} \bar{\alpha}_{\mathcal{O}}}\right] \\
& +\frac{c}{6} \log \left[\left|z_{X_1}-z_{X_2}\right|^2\right]-\frac{c}{6} \log \left[\left|1-z_{c, i}\right|^2\right]-\frac{c}{3} \log \left(\frac{2 \pi}{L}\right) .
\end{aligned}
\ee
Lastly, we proceed with the analytic continuation, $\tau=it$. 

For general cross ratios, we will calculate the entanglement entropy as the geodesic in three dimensional Euclidean anti-de sitter space ($3$d Euclidean AdS). 
In three dimensional case, the geometry is described by the so-called Bañados's metric \cite{Ban_ados_1999}, which depends only on the expectation value of the energy-momentum tensor.
The geodesic length is given as a function of the expectation value of the energy-momentum tensor.
Then,  by performing the analytic continuation as $\tau =it$, we can investigate the time dependence of the entanglement entropy as that of the analytic continued geodesic length.
\subsection{Entanglement Entropy as Geodesic Length \label{sec:EE-Geodesics}}
Consider the three dimensional Euclidean Einstein-Hilbert action with a negative cosmological constant. 
The most general vacuum solution of Einstein's equation derived from this action \cite{Ban_ados_1999} is 

\be\label{eq:metric-banados}
d s^2=\frac{L}{2} d z^2+\frac{\bar{L}}{2} d \bar{z}^2+\left(\frac{1}{y^2}+\frac{y^2}{4} L \bar{L}\right) d z d \bar{z}+\frac{d y^2}{y^2},
\ee
Here, we assume that the asymptotic geometry of this solution is three dimensional anti-de sitter space (AdS$_3$). The coordinate $y$ denotes the radial direction in the $\mathrm{AdS}_3$ bulk geometry with the conformal boundary located in the limit $y \rightarrow 0$. The coordinates $(z, \bar{z})$ are complex variables that parametrize the two-dimensional boundary where the dual CFT is defined. The functions, $L$ and $\bar{L}$, which will be specified in $(\ref{eq:LandT})$, represent the chiral and anti-chiral components of the energy-momentum tensor expectation values in the boundary CFT. These functions determine the boundary conditions of the $\mathrm{AdS}_3$ geometry through the AdS/CFT correspondence. 
The three dimensional geometry, possessing the asymptotically AdS$_3$, can be locally mapped to the pure AdS$_3$ \cite{1986CMaPh.104..207B}. 
We start with the pure AdS$_3$ geometry, and then show how it is mapped to (\ref{eq:metric-banados}).
The metric of pure AdS$_3$ is given by 
\begin{equation} \label{eq:pure-ads-3}
d s^2=\frac{d w d \overline{w}+d u^2}{u^2}.
\end{equation}
The map from 
(\ref{eq:pure-ads-3}) to (\ref{eq:metric-banados})  is defined as

\be\label{TS:BNA-ADS}
\begin{split}
&
w=f(z)- \frac{2 y^2 f^{\prime 2}(z) \overline{f}^{\prime \prime}(\overline{z}) }{4 f^{\prime}(z)\overline{f}^{\prime}(\overline{z}) +y^2  \overline{f}^{\prime \prime}(\overline{z}) f^{\prime \prime}(z)  }, \overline{w}=\overline{f}(\overline{z})- \frac{2 y^2 \overline{f}^{\prime 2}(\overline{z}) f^{\prime \prime}(z) }{4 f^{\prime}(z)\overline{f}^{\prime}(\overline{z}) +y^2  \overline{f}^{\prime \prime}(\overline{z}) f^{\prime \prime}(z)  }\\
&u=\frac{4 y \left[ f^{\prime}(z) \overline{f}^{\prime} (\overline{z})  \right]^{\frac{3}{2}}}{4 f^{\prime}(z)\overline{f}^{\prime}(\overline{z}) +y^2  \overline{f}^{\prime \prime}(\overline{z}) f^{\prime \prime}(z) },
\end{split}
\ee
where $f(z)$ and $\overline{f}(\overline{z})$ are holomorphic and anti-holomorphic functions, $f'(z)$ and $f''(z)$ are defined as $f'(z)=df(z)/dz$ and $f''(z)=d^2f(z)/dz^2$, and $\overline{f}'(\overline{z})$ and $\overline{f}''(\overline{z})$ are complex conjugate of $f'(z)$ and $f''(z)$. Note that $f(z)$ and $\bar{f}(\bar{z})$ characterize the conformal map of the boundary geometry to the plane.  The corresponding Schwarzian derivatives yield the expectation values of the stress tensor components, respectively, as
\be
\label{eq:schwarzian}
\begin{split}
    &L=\left\{f, z\right\}=\frac{3 f^{\prime \prime 2}-2 f^{\prime} f^{\prime \prime \prime}}{2 f^{\prime 2}},\\
 &\overline{L}=\left\{\overline{f}, \overline{z}\right\}=\frac{3 \overline{f}^{\prime \prime 2}-2 \overline{f}^{\prime} \overline{f}^{\prime \prime \prime}}{2 \overline{f}^{\prime 2}}.\\
\end{split}
\ee

On the other hand, as in \cite{Ban_ados_1999}, $L$ and $\overline{L}$ are determined by the  expectation value of the chiral and anti-chiral stress tensors in the dual CFT, $T(z)$ and $\overline{T}(\overline{z})$, as

\begin{equation}\label{eq:LandT}
L=\frac{12  T(z)}{c}=\frac{12   h_{\mathcal{O}}}{c 
 z^2}, \quad \overline{L}=\frac{ 12  \overline{T}(\bar{z})}{c}=\frac{12  \overline{h}_{\mathcal{O}}}{c  \bar{z}^2},
\end{equation}
where $c$, $h_{\mathcal{O}}$, and $\overline{h}_{\mathcal{O}}$ are central charge, chiral, and anti-chiral scaling dimension.
By combining (\ref{eq:schwarzian}) and (\ref{eq:LandT}), we can obtain the relation between $f(z)$ ($\overline{f}(\overline{z})$) and $T(z)$ ($\overline{T}(\overline{z}))$. We will explain the detail of this relation later.

 Then, based on AdS/CFT correspondence \cite{1999IJTP...38.1113M}, we calculate the entanglement entropy associated with the geometry in (\ref{eq:metric-banados}). 
Holographically, the entanglement entropy is determined by the area of minimal surface anchored on the subsystems \cite{2006PhRvL..96r1602R,2006JHEP...08..045R,2013JHEP...08..090L}.
Since the area of the minimal surface is invariant under the coordinate transformation, we can compute it as the geodesic length in the pure AdS$_3$ geometry in (\ref{eq:pure-ads-3}).
Let us assume that the locations of edges of subsystem are $(w_1,\overline{w}_1, \Delta_1)$ and $(w_2,\overline{w}_2, \Delta_2)$, where $\Delta_i$ is the location of the edges along the $u$-direction. 
Let $A$ denote the subsystem under consideration.
Then, the holographic entanglement entropy for $A$ can be obtained as

\be
S_A=\frac{c}{12} \log \left[\frac{\left(\overline{w}_{2}-\overline{w}_{1}\right)^2\left(w_{2}-w_{1}\right)^2}{\Delta_2^2 \Delta_1^2}\right].
\ee

In terms of $(z,\overline{z},y)$, the coordinate for (\ref{eq:metric-banados}), the holographic entanglement entropy is expressed as 
\begin{equation}\label{eq:geodesic-length}
S_{A}\left(z_1, z_2\right)=\frac{c}{12} \log \left[\frac{4\left[f\left(z_{2}\right)-f\left(z_{1}\right)\right]^2\left[\bar{f}\left(\bar{z}_{2}\right)-\bar{f}\left(\bar{z}_{1}\right)\right]^2}{f^{\prime}\left(z_{2}\right) f^{\prime}\left(z_{1}\right) \bar{f}^{\prime}\left(\bar{z}_{2}\right) \bar{f}^{\prime}\left(\bar{z}_{1}\right) \delta_1^2 \delta_2^2}\right],
\end{equation}
where $\delta_i$ denotes the location of the edges along the $y$-direction.
\subsection{Relation between Entanglement Entropy and Energy Density \label{sec:Relation-between-EE-and-energy}}
Now, we explore the relation between the holographic entanglement entropy and the expectation value of the chiral and anti-chiral stress energy tensors in the dual CFT. 
We begin by considering the relation between $f(z)$ and $T(z)$. 
By combining (\ref{eq:schwarzian}) and (\ref{eq:LandT}), we obtain the equation for $f(z)$ and $T(z)$ as
\be\label{eq:ODE-f-T}
-\f{12 }{c} T(z)= \frac{f^{\prime \prime \prime}(z) f^{\prime}(z) - \frac{3}{2} f^{\prime \prime}(z)^2}{f^{\prime}(z)^2}.
\ee

Define a new holomorphic function $g(z)$ as $g(z)=f'(z)$, and then (\ref{eq:ODE-f-T}) reduces to a second-order differential equation,
\be\label{eq:gt}
-\f{12 }{c} T(z)= \frac{g^{\prime \prime}(z) g(z) - \frac{3}{2} g^{\prime}(z)^2}{g(z)^2}.
\ee
Then, define a new holomorphic function, $p(z)$, as $p(z)=\f{g'(z)}{g(z)}$. This new function must be subject to the following equation,
\be
p'(z)=\left(\f{g'(z)}{g(z)}\right)'=\f{g''(z)g(z)-g'(z)^2}{g(z)^2}.
\ee
Consequently, (\ref{eq:gt}) reduces to a first-order ordinary differential equation,
\be\label{eq:pt}
-\f{12 }{c} T(z)= p'(z)-\f{1}{2}p(z)^2.
\ee
This is equivalent to a Riccati equation,
\begin{equation} \label{eq:Riccati-1}
p^{\prime}(z)=q_0(z)+q_1(z) p(z)+q_2(z) p^2(z),
\end{equation}
with $q_0(z)=-\f{12 \pi}{c}T(z)$, $q_1(z)=0$, and $q_2(z)=\f{1}{2}$.
Then, we introduce a new function of $z$ as $v(z)=p(z) q_2(z)$, so that (\ref{eq:Riccati-1}) reduces to 
\be \label{eq:Riccati-2}
v(z)^{\prime}=v(z)^2+R(z) v(z)+S(z),
\ee
where the two functions of $z$, $S(z)$ and $R(z)$, are given as $S(z)=q_2(z) q_0(z)=-\f{6 }{c} T(z)$ and $R(z)=q_1(z)+\frac{q_2^{\prime}(z)}{q_2(z)}=0$.
Define a function of $z$, $u(z)$, as $v(z)=-\frac{u^{\prime}(z)}{u(z)}$ and substitute it to (\ref{eq:Riccati-2}). 
Then, (\ref{eq:Riccati-2}) results in the linear second-order ordinary differential equation of $u(z)$,
\be\label{eq:Riccati-3}
u^{\prime \prime}-R(z) u^{\prime}+S(z) u=0.
\ee
Let $u_1(z)$ and $u_2(z)$ be two linearly independent solutions to this equation. Then, the general solution to (\ref{eq:Riccati-3}) can be written as
\be\label{eq:Riccati-3-solution}
u(z)=c_1 u_1(z)+c_2 u_2(z),
\ee
where $c_1$ and $c_2$ are integration constants. Correspondingly, the general solution of (\ref{eq:pt}) is given by
\begin{equation} \label{eq:pt-solution}
	p(z) = -2 \left( \log \left( c_1 u_1(z) + c_2 u_2(z) \right) \right)'.
\end{equation}
Furthermore, since \( p(z) = \left( \log f'(z) \right)' \), we can integrate both sides to obtain
\begin{equation} \label{eq:log-f-prime}
	\log f'(z) = -2 \log \left( c_1 u_1(z) + c_2 u_2(z) \right) + \text{const.}
\end{equation}
Absorbing the additive constant into \( c_1 \) and \( c_2 \), we arrive at
\begin{equation} \label{eq:f-prime-u}
	f'(z) = \left( c_1 u_1(z) + c_2 u_2(z) \right)^{-2},
\end{equation}
\begin{equation} \label{eq:f-from-u}
	f(z) = \int^z_{z_0} \left( c_1 u_1(\xi) + c_2 u_2(\xi) \right)^{-2} d\xi ,
\end{equation}
where \( z_0 \) is a reference point. The integration is performed along the $z$-direction, as a formal inverse operation of the holomorphic derivative.
This expression provides an explicit reconstruction of the conformal map, \( f(z) \), from the energy-momentum tensor, \( T(z) \), via the solutions (\ref{eq:Riccati-3-solution}) of a second-order linear ordinary differential equation.
We define a function of \(z\), \(\mathcal{T}(z)\), as
\be
\mathcal{T}(z)=\mathcal{T}(T(z)):=\f{f'(z)}{f(z)}=\f{\left(c_1 u_1(z)+c_2 u_2(z)\right)^{-2}}{\int^{z}_{z_0} \left(c_1 u_1(z)+c_2 u_2(z)\right)^{-2} dz}=\f{u(z)^{-2}}{\int^{z}_{z_0}  u(z)^{-2} dz},
\ee 
where $u_1$ and $u_2$ are two linearly independent solutions of \be
\begin{aligned} \label{eq:relation-u-and-Tz}
\frac{u^{\prime \prime}}{u} =\f{6}{c} T(z)
\end{aligned}.
\ee
Thus, $\mathcal{T}$ is implicitly determined by $T(z)$ through (\ref{eq:relation-u-and-Tz}). Here, we can determine the three  previously undetermined parameters, $c_1$, $c_2$, and $z_0$ , by specifying a concrete setup.

Let us go back to (\ref{eq:geodesic-length}), we can rewrite it as
\begin{equation}\label{eq:mid-EE-energy}
\begin{split}
S_{A}\left(z_1, z_2\right)&=\frac{c}{12} \log \left[\frac{4\f{\left[f\left(z_{2}\right)-f\left(z_{1}\right)\right]^2}{f\left(z_{2}\right)f\left(z_{1}\right)}\f{\left[\bar{f}\left(\bar{z}_{2}\right)-\bar{f}\left(\bar{z}_{1}\right)\right]^2}{\bar{f}\left(\bar{z}_{2}\right)\bar{f}\left(\bar{z}_{1}\right)}}{\f{f^{\prime}\left(z_{2}\right) f^{\prime}\left(z_{1}\right) \bar{f}^{\prime}\left(\bar{z}_{2}\right) \bar{f}^{\prime}\left(\bar{z}_{1}\right) }{f\left(z_{2}\right)f\left(z_{1}\right) \bar{f}\left(\bar{z}_{2}\right)\bar{f}\left(\bar{z}_{1}\right)}\delta_1^2 \delta_2^2}\right]\\
&=\frac{c}{12} \log \left[\frac{4\left[\left(\frac{f\left(z_{2}\right)}{f\left(z_{1}\right)}\right)^{\frac{1}{2}}-\left(\frac{f\left(z_{1}\right)}{f\left(z_{2}\right)}\right)^{\frac{1}{2}}\right]^2\left[\left(\frac{\bar{f}\left(\bar{z}_{2}\right)}{\bar{f}\left(\bar{z}_{1}\right)}\right)^{\frac{1}{2}}-\left(\frac{\bar{f}\left(\bar{z}_{1}\right)}{\bar{f}\left(\bar{z}_{2}\right)}\right)^{\frac{1}{2}}\right]^2}{\frac{f^{\prime}\left(z_{2}\right) f^{\prime}\left(z_{1}\right) \bar{f}^{\prime}\left(\bar{z}_{2}\right) \bar{f}^{\prime}\left(\bar{z}_{1}\right) }{f\left(z_{2}\right)f\left(z_{1}\right) \bar{f}\left(\bar{z}_{2}\right)\bar{f}\left(\bar{z}_{1}\right)}\delta_1^2 \delta_2^2}\right].\\
\end{split}
\end{equation}
Note that $\frac{d \log \left[f(z)\right] }{d z}= \frac{f'(z)}{f(z)}$, and therefore, $\int_{z_1}^{z_2} \frac{f'(z)}{f(z)} dz = \log \left[f(z_2)\right] -\log \left[f(z_1)\right] =\log\left[\frac{f(z_2)}{f(z_1)}\right]$.
Then, simplify the first term in the numerator of the expression on the right-hand side of the second equality in (\ref{eq:mid-EE-energy}) as
\be\small \label{eq:exponential-function-of-fz}
\left(\frac{f\left(z_2\right)}{f\left(z_1\right)}\right)^{\frac{1}{2}}-\left(\frac{f\left(z_1\right)}{f\left(z_2\right)}\right)^{\frac{1}{2}}= \exp{\left(\f{1}{2} \int_{z_1}^{z_2} \frac{f^{\prime}(z)}{f(z)} d z\right)}- \exp{\left(-\f{1}{2} \int_{z_1}^{z_2} \frac{f^{\prime}(z)}{f(z)} d z\right)}=2  \sinh{\left(\frac{1}{2} \int_{z_1}^{z_2} \frac{f^{\prime}(z)}{f(z)} d z\right)}.
\ee
The anti-chiral part of the energy-momentum density has the counterparts of (\ref{eq:exponential-function-of-fz}) and (\ref{eq:geodesic-length}), and we can define that of $\overline{\mathcal{T}}(\overline{z})$ as
\be
\overline{\mathcal{T}}(\overline{z})=\overline{\mathcal{T}}(T(\overline{z})):=\f{\overline{f}^{\prime}(\overline{z})}{\overline{f}(\overline{z})}.
\ee
Substitute (\ref{eq:exponential-function-of-fz}) into (\ref{eq:geodesic-length}), and then exploit the identity, $\mathcal{T}(z):=\frac{f^{\prime}(z)}{f(z)}$.
Then, for the anti-chiral part of the energy-momentum tensor, do the same calculation as the chiral one. 
Consequently, the entanglement entropy results in 
\begin{equation}  \label{eq:general-entanglement-entropy}
\begin{split}
S_A(z_1,z_2) = \min \Bigg\{ 
&\frac{c}{6} \log \left[ \frac{8 \left| \sinh \left( \frac{1}{2} \int_A \mathcal{T}(z) dz \right) \right| }{\delta^2 \left[\mathcal{T}(z_1) \overline{\mathcal{T}}(\overline{z}_1)\right]^{1/2} \left[\mathcal{T}(z_2) \overline{\mathcal{T}}(\overline{z}_2)\right]^{1/2}} \right], \\
&\frac{c}{6} \log \left[ \frac{8 \left| \sinh \left( \frac{1}{2} \int_{\overline{A}} \mathcal{T}(z) dz \right) \right| }{\delta^2 \left[\mathcal{T}(z_1) \overline{\mathcal{T}}(\overline{z}_1)\right]^{1/2} \left[\mathcal{T}(z_2) \overline{\mathcal{T}}(\overline{z}_2)\right]^{1/2}} \right]
\Bigg\},
\end{split}
\end{equation}
where for simplicity, we set $\delta_1=\delta_2 =\delta$.

Let us consider the expression for the holographic entanglement entropy for the system considered in this paper. 
For the chiral and anti-chiral energy-momentum densities induced by the insertion of the local operator, $f(z)$ and $\overline{f}(\overline{z})$ satisfy the following equation,
\be
\mathcal{T}(z)=\f{f'(z)}{f(z)}=C \sqrt{T(z)},~ \overline{\mathcal{T}}(\overline{z})=\f{\overline{f}'(\overline{z})}{\overline{f}(\overline{z})}=C^{*} \sqrt{\overline{T}(\overline{z})},
\ee
where $\cdot^{*}$ denotes the complex conjugate of $\cdot$.
In this case, the holographic entanglement entropy is simplified as 
\begin{equation}\label{eq:entanglement-entropy-and-energy-density}
\begin{split}
S_A(z_1, z_2) = \min \Bigg\{ \,
& \frac{c}{6} \log \left[ \frac{8 \left| \sinh \left( \frac{C}{2} \int_A \sqrt{T(z)} dz \right) \right| }{ \delta^2 |C|^2 \left[ T(z_1)\overline{T}(\overline{z}_1) \right]^{1/4} \left[ T(z_2)\overline{T}(\overline{z}_2) \right]^{1/4} } \right], \\
& \frac{c}{6} \log \left[ \frac{8 \left| \sinh \left( \frac{C}{2} \int_{\overline{A}} \sqrt{T(z)} dz \right) \right| }{ \delta^2 |C|^2 \left[ T(z_1)\overline{T}(\overline{z}_1) \right]^{1/4} \left[ T(z_2)\overline{T}(\overline{z}_2) \right]^{1/4} } \right]
\, \Bigg\}.
\end{split}
\end{equation}
In the system considered in this paper, $C$, $T(z)$, and $\overline{T}(\overline{z})$ are given by 
\be
T(z)=h_{\mathcal{O}}
\left[
\frac{1}{z - z^{\mathrm{new},i}_{\epsilon}} - \frac{1}{z - z^{\mathrm{new},i}_{-\epsilon}}
\right]^2, ~\overline{T}(\overline{z})=\bar{h}_{\mathcal{O}}
\left[
\frac{1}{\bar z - \bar z^{\mathrm{new},i}_{\epsilon}} - \frac{1}{\bar z - \bar z^{\mathrm{new},i}_{-\epsilon}}
\right]^2,  C=-\f{\sqrt{1-24 h_{\mathcal{O}} / c}}{\sqrt{h_{\mathcal{O}}}},
\ee
where for $h_{\mathcal{O}}<\f{c}{24}$, $C$ is real, while, $h_{\mathcal{O}}>\f{c}{24}$, $C=-i \f{\sqrt{24 h_{\mathcal{O}} / c -1}}{\sqrt{h_{\mathcal{O}}}}$ is a purely imaginary number.
The special case, $h_{\mathcal{O}}=\f{c}{24}$, will be discussed in appendix \ref{ch:special}. 
It should be noted that the multivaluedness of the geodesic expression is inherent in the process of analytic continuation from Euclidean spacetime to Minkowski spacetime. During this process, the energy-momentum tensor $T(z)$ and its conjugate energy-momentum tensor $\overline{T}(\bar{z})$ are no longer complex conjugates of each other. Therefore, the multivaluedness with respect to $z$ and $\bar{z}$ still needs to be considered, and geodesics of different lengths are included in this single expression.
\subsubsection{Energy density}
Here, our focus shifts to investigating the temporal evolution of the expectation value of the chiral and anti-chiral energy momentum tensors. 
For our analysis, we begin by computing the energy density within the Euclidean path integral framework and subsequently perform the analytical continuation to real-time dynamics.
We define the expectation values of the chiral and anti-chiral components of the energy-momentum tensor as
\begin{equation}
	\begin{split}
		\bigl\langle T_{ww}(w_X)\bigr\rangle_{i,E}
		&= tr\!\bigl(\rho_{i,E}\,T_{ww}(w_X)\bigr) \\
		&= \frac{\bigl\langle
			\tilde{\mathcal{O}}^{\text{Euc.}}_{H,n,i;-}\!\bigl(
			w^{\text{new},i}_{\epsilon},
			\overline{w}^{\text{new},i}_{\epsilon}\bigr)\,
			T_{ww}(w_X)\,
			\mathcal{O}^{\text{Euc.}}_{H,n,i;+}\!\bigl(
			w^{\text{new},i}_{-\epsilon},
			\overline{w}^{\text{new},i}_{-\epsilon}\bigr)
			\bigr\rangle}{
			\bigl\langle
			\tilde{\mathcal{O}}^{\text{Euc.}}_{H,n,i;-}\!\bigl(
			w^{\text{new},i}_{\epsilon},
			\overline{w}^{\text{new},i}_{\epsilon}\bigr)\,
			\mathcal{O}^{\text{Euc.}}_{H,n,i;+}\!\bigl(
			w^{\text{new},i}_{-\epsilon},
			\overline{w}^{\text{new},i}_{-\epsilon}\bigr)
			\bigr\rangle},\\[6pt]
		\bigl\langle T_{\overline{w}\overline{w}}(\overline{w}_X)\bigr\rangle_{i,E}
		&=tr\!\bigl(\rho_{i,E}\,
		T_{\overline{w}\overline{w}}(\overline{w}_X)\bigr) \\
		&= \frac{\bigl\langle
			\tilde{\mathcal{O}}^{\text{Euc.}}_{H,n,i;-}\!\bigl(
			w^{\text{new},i}_{\epsilon},
			\overline{w}^{\text{new},i}_{\epsilon}\bigr)\,
			T_{\overline{w}\overline{w}}(\overline{w}_X)\,
			\mathcal{O}^{\text{Euc.}}_{H,n,i;+}\!\bigl(
			w^{\text{new},i}_{-\epsilon},
			\overline{w}^{\text{new},i}_{-\epsilon}\bigr)
			\bigr\rangle}{
			\bigl\langle
			\tilde{\mathcal{O}}^{\text{Euc.}}_{H,n,i;-}\!\bigl(
			w^{\text{new},i}_{\epsilon},
			\overline{w}^{\text{new},i}_{\epsilon}\bigr)\,
			\mathcal{O}^{\text{Euc.}}_{H,n,i;+}\!\bigl(
			w^{\text{new},i}_{-\epsilon},
			\overline{w}^{\text{new},i}_{-\epsilon}\bigr)
			\bigr\rangle},
	\end{split}
\end{equation}
where the energy density is assumed to be defined as the linear combination of the chiral and anti-chiral parts of the energy density, $T=T_{ww}+T_{\overline{w}\overline{w}}$. 
We call $\left\langle T_{ww} \right \rangle_i$ and $\left\langle T_{\overline{w}\overline{w}}\right \rangle_i$ chiral and anti-chiral energy densities, respectively.
By performing the conformal map from the cylinder to the plane and using the Ward-Takahashi identity, we obtain the expectation values of chiral and anti-chiral energy densities as \footnote{Here, the Ward-Takahashi identity is \be\begin{split}
    &\left\langle T_{zz}(z) \mathcal{O}(z_1,\overline{z}_1)\mathcal{O}(z_2,\overline{z}_2) \right \rangle=\sum_{i=1}^2\left[\f{h_{\mathcal{O}}}{\left(z-z_i\right)^2}+\f{1}{z-z_i}\partial_{z_i}\right]\left\langle \mathcal{O}(z_1,\overline{z}_1)\mathcal{O}(z_2,\overline{z}_2) \right \rangle,\\
    &\left\langle T_{\overline{z}\overline{z}}(\overline{z}) \mathcal{O}(z_1,\overline{z}_1)\mathcal{O}(z_2,\overline{z}_2) \right \rangle=\sum_{i=1}^2\left[\f{h_{\mathcal{O}}}{\left(\overline{z}-\overline{z}_i\right)^2}+\f{1}{z-z_i}\partial_{\overline{z}_i}\right]\left\langle \mathcal{O}(z_1,\overline{z}_1)\mathcal{O}(z_2,\overline{z}_2) \right \rangle.\\
\end{split}\ee}
\begin{equation}
\begin{split}
    \left \langle T_{ww} (w_X,\overline{w}_X)\right \rangle_{i,E} &=-\frac{c}{24}\left(\frac{2\pi}{L}\right)^2+h_{\mathcal{O}}\left(\frac{dz_{X}}{dw_X}\right)^2\left[\frac{1}{z_X - z^{\text{new},i}_{\epsilon}} - \frac{1}{z_X - z^{\text{new},i}_{-\epsilon}}\right]^2,\\
    \left \langle T_{\overline{w}\overline{w}} (w_X,\overline{w}_X)\right \rangle_{i,E} &=-\frac{c}{24}\left(\frac{2\pi}{L}\right)^2+h_{\mathcal{O}}\left(\frac{d\overline{z}_{X}}{d\overline{w}_X}\right)^2\left[\frac{1}{\overline{z}_X - \overline{z}^{\text{new},i}_{\epsilon}} - \frac{1}{\overline{z}_X - \overline{z}^{\text{new},i}_{-\epsilon}}\right]^2.\\
\end{split}
\end{equation}

\begin{figure}[htbp!]
	\centering
	\subfloat[$c=1, h_{\mathcal{O}}=1, \epsilon=50, L=10000, x=2500
	.$]{\includegraphics[width=.4\columnwidth]{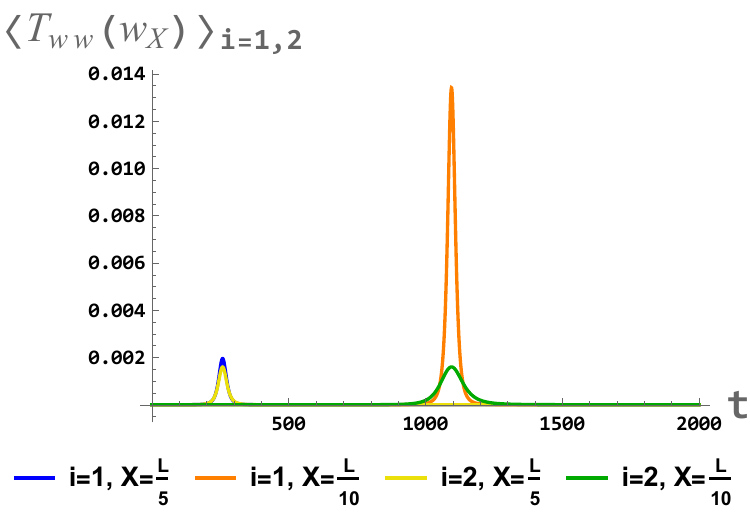}}\hspace{5pt}
	\subfloat[$c=1, h_{\mathcal{O}}=1, \epsilon=50, L=10000, x=2500
	.$]{\includegraphics[width=.4\columnwidth]{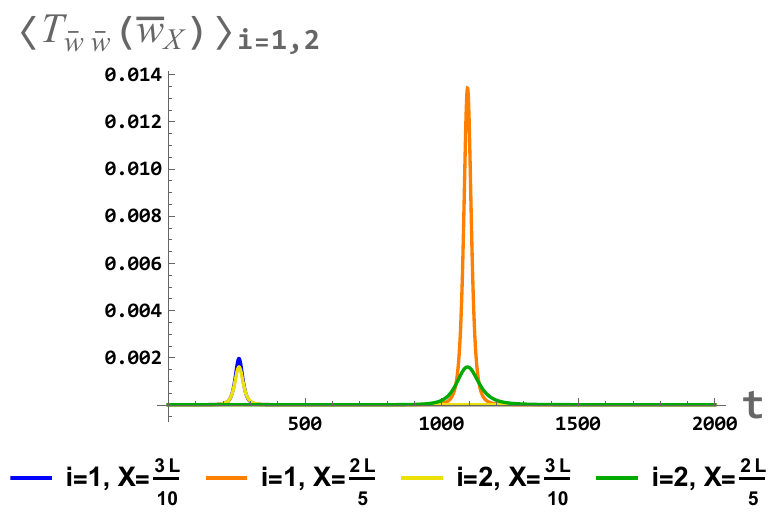}}\\
	\subfloat[$c=1, h_{\mathcal{O}}=1,\epsilon_1=50, \epsilon_2=20, L=10000, x=2500.$]{\includegraphics[width=.4\columnwidth]{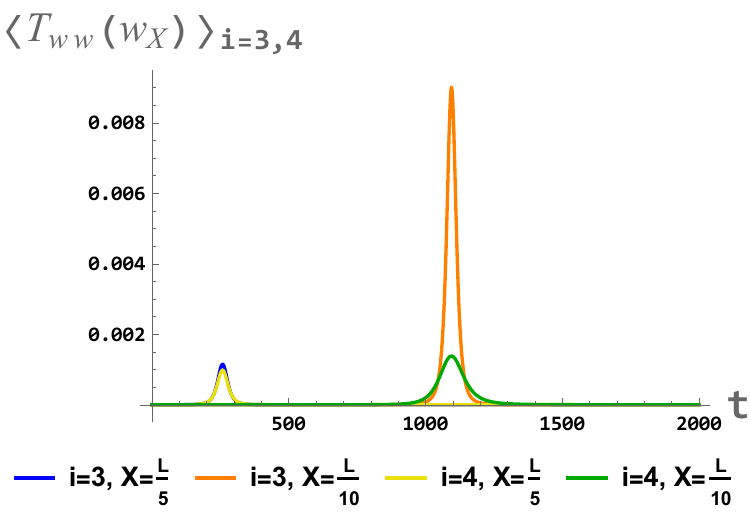}}\hspace{5pt}
	\subfloat[$c=1, h_{\mathcal{O}}=1,\epsilon_1=50, \epsilon_2=20, L=10000, x=2500.$]{\includegraphics[width=.4\columnwidth]{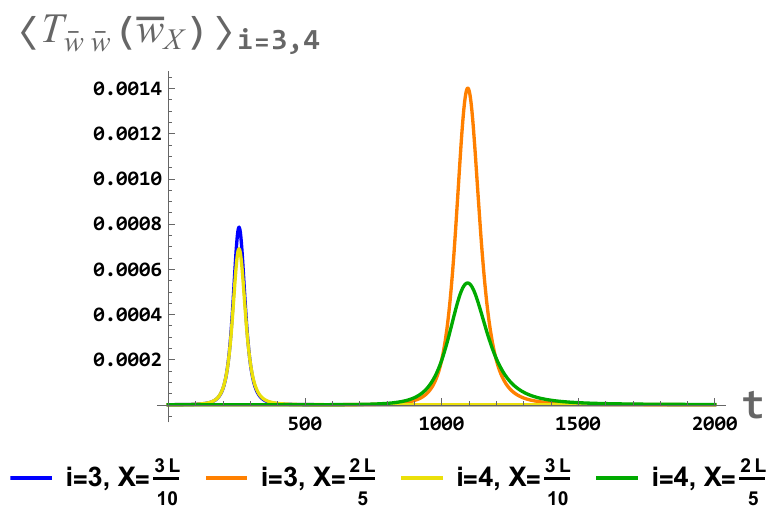}}
	\caption{The chiral and anti-chiral energy densities for $i=1 \sim 4$, as functions of $t$. Here, we insert the local operator into $x=\frac{L}{4}$. In panels (I) and (II), we show the time dependence of the chiral and anti-chiral energy densities for $i=1,2$, while in (III) and (IV), we show the time dependence of the chiral and anti-chiral energy densities for $i=3,4$. The panels (I) and (III) show the time dependence of the chiral energy density, while (II) and (IV) show that of the anti-chiral one.}\label{Fig:ED}
\end{figure}

We present the plot of chiral and anti-chiral energy densities without the small $\epsilon$ expansion in Fig.\ref{Fig:ED}. 
In (I)-(IV) of Fig. \ref{Fig:ED}, we show the chiral and anti-chiral energy densities, as the functions of $t$ . The panels, (I) and (II), show that the chiral and anti-chiral energy densities have a peak at certain times. These times are consistent with the quasiparticle picture. 
In this picture, two local excitations, called quasiparticles, emerge at the insertion point of the local operator at $t=0$. They propagate to positive and negative directions of $x$ with the speed of the massless particles moving on the geometry where Hamiltonian, for Lorentzian time evolution, is defined. 
	We call the particles, moving to the positive and negative direction of $x$, right and left moving particles, respectively. 
	For $H_{\text{2-SSD}}$, the curved background is given by
	\be
	ds^2=-4\sin^4{\left(\f{2\pi x}{L}\right)}dt^2+dx^2.
	\ee
	Then, the velocities of the massless particles are given by $v= \pm 2\sin^2{\left(\f{2\pi x}{L}\right)}$.
	The time for the peak of the energy density to emerge at position $X$ corresponds to that for the quasiparticles induced by the local operator to arrive at that point.
	 The panels, (III) and (IV), show how $e^{-\epsilon H_{\text{SSD}}}$ affects the time dependence of energy densities.  
	In the spatial region, $0\le X\le\f{L}{2}$, the value of energy density peak decreases as $X$ increases. 
	This may be due to the effect of $e^{-\epsilon H_{\text{SSD}}}$ on the energy densities.
	When $X$ becomes larger in the spatial region, $0\le X\le\f{L}{2}$, the effect of the inhomogeneous Hamiltonian, $H_{\text{SSD}}$, may become larger, and suppress the energy excitations in that region.
    We can also define the expectation values of the chiral and anti-chiral energy-momentum tensors  in $(z, \bar{z})$ as 
\be\label{eq:energy-density-z}
\begin{split}
	\left\langle T\left(z_{X}\right)\right\rangle_{i,E}
	&= \Tr\left( \rho_{i,E}\,T\left(z_{X}\right) \right) = \text{\footnotesize{$\displaystyle
			\frac{
				\left\langle
				\tilde{\mathcal{O}}^{\text{Euc.}}_{H,n,i;-}\left(
				z^{\text{new},i}_{\epsilon},
				\,\bar z^{\text{new},i}_{\epsilon}
				\right)
				\,T\left(z_{X}\right)\,
				\mathcal{O}^{\text{Euc.}}_{H,n,i;+}\left(
				z^{\text{new},i}_{-\epsilon},
				\,\bar z^{\text{new},i}_{-\epsilon}
				\right)
				\right\rangle}{
				\left\langle
				\tilde{\mathcal{O}}^{\text{Euc.}}_{H,n,i;-}\left(
				z^{\text{new},i}_{\epsilon},
				\,\bar z^{\text{new},i}_{\epsilon}
				\right)
				\mathcal{O}^{\text{Euc.}}_{H,n,i;+}\left(
				z^{\text{new},i}_{-\epsilon},
				\,\bar z^{\text{new},i}_{-\epsilon}
				\right)
				\right\rangle}$}},\\[6pt]
	\left\langle \bar T\left( \bar z_{X} \right) \right\rangle_{i,E}
	&= \Tr\left( \rho_{i,E}\,\bar T\left( \bar z_{X} \right) \right) = \text{\footnotesize{$\displaystyle
			\frac{
				\left\langle
				\tilde{\mathcal{O}}^{\text{Euc.}}_{H,n,i;-}\left(
				z^{\text{new},i}_{\epsilon},
				\,\bar z^{\text{new},i}_{\epsilon}
				\right)
				\,\bar T\left( \bar z_{X} \right)\,
				\mathcal{O}^{\text{Euc.}}_{H,n,i;+}\left(
				z^{\text{new},i}_{-\epsilon},
				\,\bar z^{\text{new},i}_{-\epsilon}
				\right)
				\right\rangle}{
				\left\langle
				\tilde{\mathcal{O}}^{\text{Euc.}}_{H,n,i;-}\left(
				z^{\text{new},i}_{\epsilon},
				\,\bar z^{\text{new},i}_{\epsilon}
				\right)
				\mathcal{O}^{\text{Euc.}}_{H,n,i;+}\left(
				z^{\text{new},i}_{-\epsilon},
				\,\bar z^{\text{new},i}_{-\epsilon}
				\right)
				\right\rangle}$}}.
\end{split}
\ee

\subsection{The Effect of the Time Ordering\label{sec:Warm-up}}
\begin{figure}[htbp]
\centering
\subfloat[case a]{\includegraphics[width=.45\columnwidth]{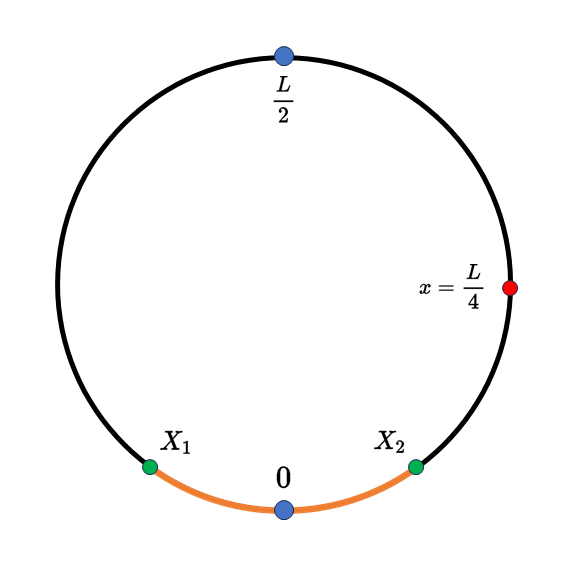}}\hspace{5pt}
\subfloat[case b]
{\includegraphics[width=.45\columnwidth]{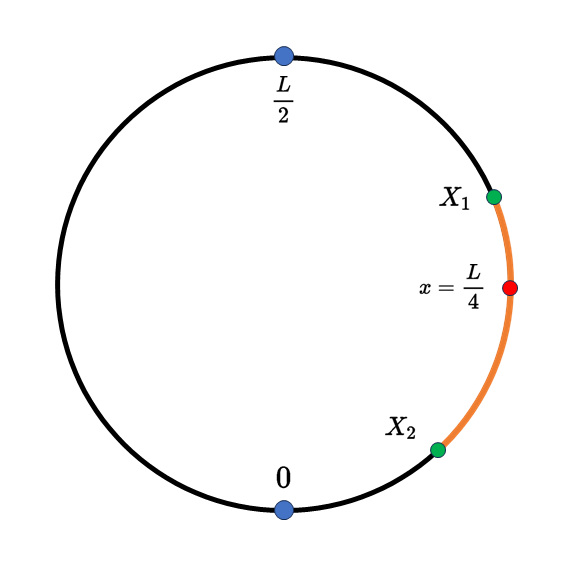}}\hspace{5pt}
	\caption{ The subsystems under consideration. In the left panel, we show the subsystem containing a fixed point, $x=0$.
    In the right panel, we show the one containing the insertion point of the local operator though it does not contain any fixed points.}\label{Fig:case-ab}
\end{figure}

Now, move on to the first main research topic of this paper.
To investigate the time ordering effect on the entanglement dynamics, we begin by calculating the entanglement entropy for the setups labeled by \( i = 1,2 \).
For simplicity, we insert the local operator at $x=\f{L}{4}$, and then consider the following subsystems (see Fig. \ref{Fig:case-ab}):
\be \label{eq:def-case-ab}
\begin{split}
   &\text{Case a:}~ A=[X_1,L]\cup [0,X_2],\\
    &\text{Case b:}~ A=[X_2,X_1], 0<X_2<\f{L}{4}<X_1, X_1-\f{L}{4}<\f{L}{4}-X_2.\\
\end{split}
\ee
We perform the analytic continuation, and then take the small $\epsilon$ expansion.
By the second order in the small $\epsilon$ expansion, the analytically continued cross ratios are given by 
\be
\begin{aligned}
& z_{c, i} \approx 1+i \epsilon f_i\left(t\right), \\
& \bar{z}_{c, i} \approx 1+i \epsilon g_i\left(t\right),\\
\end{aligned}
\ee
where the functions of $t$, $f_i(t)$ and $g_i(t)$, are defined as
\be
\begin{aligned}
& f_1(t)=-\frac{2  \pi L \sin \left(\frac{\pi  \left(X_1-X_2\right)}{L}\right)}{ \left(L^2+16 \pi ^2 t^2\right) \sin \left(\phi -\frac{\pi  X_1}{L}\right) \sin \left(\phi -\frac{\pi  X_2}{L}\right)}, \\
& g_1(t)=\frac{2  \pi  L  \sin \left(\frac{\pi  \left(X_1-X_2\right)}{L}\right)}{ \left(L^2+16 \pi ^2 t^2\right) \sin \left(\psi -\frac{\pi  X_1}{L}\right) \sin \left(\psi -\frac{\pi  X_2}{L}\right)}, \\
& f_2(t)=-\frac{2  \pi    \sin \left(\frac{\pi  \left(X_1-X_2\right)}{L}\right)}{L \sin \left(\phi -\frac{\pi  X_1}{L}\right) \sin \left(\phi -\frac{\pi  X_2}{L}\right)}, \\
& g_2(t)=\frac{2  \pi    \sin \left(\frac{\pi  \left(X_1-X_2\right)}{L}\right)}{L \sin \left(\psi -\frac{\pi  X_1}{L}\right) \sin \left(\psi -\frac{\pi  X_2}{L}\right)}.
\end{aligned}
\ee
Here, the angular functions of $t$,  $\phi$ and $\psi$, are defined as
\be
\begin{split}
    e^{2 i \phi }=\frac{4 \pi  t+i L}{\sqrt{L^2+16 \pi ^2 t^2}}=\sqrt{\frac{4 \pi  t+i L}{4 \pi  t-i L}},\\
    e^{2 i \psi}=\frac{-4 \pi  t+i L}{\sqrt{L^2+16 \pi ^2 t^2}}=-\sqrt{\frac{4 \pi  t-i L}{4 \pi  t+i L}}.
\end{split}
\ee
Since for any positive time $t$, the derivative, $\frac{d\phi}{dt}=-\frac{2 \pi  L}{L^2+16 \pi ^2 t^2}$, is negative, the angular function of $t$, $\phi(t)$, monotonically decreases with $t$.
It starts from the initial value $\phi(t=0) = \frac{\pi}{4}$, and then saturates to $\phi(t=\infty) = 0$ for the large time region, $t\gg 1$. 
In contrast, the derivative, $\frac{d\psi}{dt}=\frac{2 \pi  L}{L^2+16 \pi ^2 t^2}$, is positive.
Therefore,  $\psi(t)$ monotonically increases with $t$, starting from the initial value of $\psi(t=0) = \frac{\pi}{4}$, and then saturates to $\psi(t=\infty) = \frac{\pi}{2}$ for the large time regime.

Since the Euclidean entanglement entropy in (\ref{eq:entanglement-formula-in-Euclidean}) has $(z^{\gamma}_{c,i}, \overline{z}^{\gamma}_{c,i})$, where $\gamma$ is not integer, after the analytic continuation, it becomes a multi-valued function of the cross ratios. 
Therefore, we need to take into account the analytic-continued entanglement entropies depending on the choice of the branch associated with the origin, 
\be
z_{c,i} \rightarrow e^{\pm  2i\pi }z_{c,i}, \overline{z}_{c,i} \rightarrow e^{\pm 2i \pi}\overline{z}_{c,i}.
\ee
In the following, we assume that the time dependence of the entanglement entropy is determined by the minimal one, which follows the physical picture called the quasiparticle picture, of the analytic-continued entanglement entropies \cite{2015JHEP...02..171A}.
In this quasiparticle picture, we assume that the local operator induces the entangled pair at its insertion point.
This pair is constructed of two massless particles (quasiparticles) which are entangled with each other.
One of them propagates to the left, while the other propagates to the right.
Their speeds are determined by the curved background, as derived in appendix \ref{ch:apa},
\be \label{eq:velocity-of-QP}
\left|v(x)\right|=2\sin^2{\left(\f{2\pi x}{L}\right)},
\ee
where the velocities around $x=0, \f{L}{2}$ are almost zero, $\left|v(x\approx 0,\f{L}{2})\right| \approx 0$. 
This suggests that for the large time regime, the quasiparticles will accumulate around $x=0$ or $x=\f{L}{2}$.
We assume that in the time region, where only one quasiparticle is in $A$, the subsystem under consideration, the quantum entanglement between them contributes to $S_A$.
Under these assumptions, in case a, the entanglement between quasiparticles continues to contribute to $S_A$ even for the large time regime, while in case b, it does not.
\subsubsection{case a}
\begin{figure}[htbp!]
	\centering
	\includegraphics[width=0.6\textwidth]{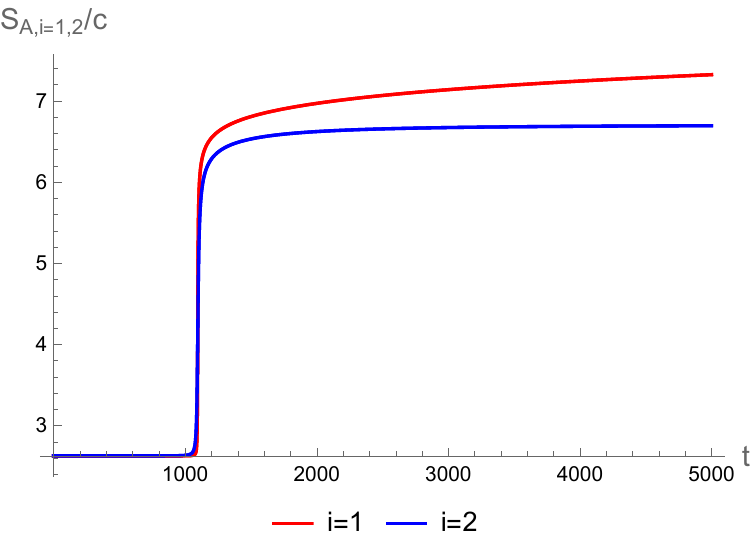} %
	\caption{
		Entanglement entropy as a function of time, with parameters 
		$L = 10000$, $c = 1000$, $h_{\mathcal{O}} = \frac{c}{24} + 10$, 
		$\delta = 1$, $x = 2500$, $X_1 = 9000$, $X_2 = 1000$, 
		and $\epsilon = 10$.
    }
	\label{fig:case_a_ee}
\end{figure}

Now, we present the time dependence of $S_A$ for case a with $i=1,2$. 
As shown in Fig.~\ref{fig:case_a_ee}, $S_A$ exhibits a characteristic growth after a certain time, reflecting the contribution from quasiparticle propagation.
In this case, one of the quasiparticles induced by the local operator propagates to $x=\f{L}{2}$, while one of them propagates to $x=0$.
Let $t_{X_2}$ denote the time for the quasiparticle induced at $x=\f{L}{4}$ to arrive at $x=X_2$,
\be
t_{X_2} = \frac{L \cot \left(\frac{2 \pi  X_2}{L}\right)}{4 \pi }.
\ee
Before $t=t_{X_2}$, both quasiparticles are in $A$, while after $t=t_{X_2}$, only one of them is in $A$.
Therefore, by the second order of the small $\epsilon$ expansion, $\epsilon \ll 1$, the time dependence of $S_{A,i=1,2}$ is given by
\be
\begin{split}
       &S_{A,1}\approx \frac{c}{3}  \log \left[\frac{L }{ \pi } \sin \left(\frac{ \pi  \left(X_1-X_2\right)}{L}\right)\right]   +  \begin{cases}
0 &  t_{X_2}>t>0\\
\frac{c}{6} \log \left[\frac{\sin \left[\pi \alpha_{\mathcal{O}}\right]}{\pi \alpha_{\mathcal{O}}}\right]-\frac{c}{6}  \log \left[-\epsilon f_1\left( t\right)\right] &   t>t_{X_2} \\
    \end{cases},\\
    &S_{A,2}\approx \frac{c}{3}  \log \left[\frac{L }{\pi } \sin \left(\frac{ \pi  \left(X_1-X_2\right)}{L}\right)\right] +  \begin{cases}
0 &  t_{X_2}>t>0\\
\frac{c}{6} \log \left[\frac{\sin \left[\pi \alpha_{\mathcal{O}}\right]}{\pi \alpha_{\mathcal{O}}}\right]-\frac{c}{6}  \log \left[-\epsilon f_2\left( t\right)\right]&   t>t_{X_2} \\
    \end{cases}.
\end{split}
\ee
Thus, before $t=t_{X_2}$, the entanglement entropy is given by that for the vacuum state in the circle with circumference of $L$ \cite{2004JSMTE..06..002C,2009JPhA...42X4005C}. 
After $t=t_{X_2}$, the quantum entanglement between quasiparticles contributes to $S_A$. 
In this time regime, the derivatives of $S_{A,i}$ with respect to $t$ are positive,
\be
\begin{split}
    \f{d S_{A,1}}{d t}=\frac{\pi  c \left(-L \cot \left(\phi -\frac{\pi  X_1}{L}\right)-L \cot \left(\phi -\frac{\pi  X_2}{L}\right)+16 \pi  t\right)}{3 \left(L^2+16 \pi ^2 t^2\right)}>0,&\\
    \f{d S_{A,2}}{d t}=-\frac{\pi  c L \left(\cot \left(\phi -\frac{\pi  X_1}{L}\right)+\cot \left(\phi -\frac{\pi  X_2}{L}\right)\right)}{3 \left(L^2+16 \pi ^2 t^2\right)} > 0.&\\
\end{split}
\ee
Thus, $S_{A,i=1,2}$ monotonically grows with $t$.
The contribution in the case $i=1$ is different from that in $i=2$.
To closely look at the significant difference between them, consider the time dependence of $S_{A,i}$ for the large time regime, that is when $t \gg L$.
In this large time region, $t \gg L$, the entanglement entropy is approximately given by 
\be
\begin{split}
S_{A,i}&\approx \frac{c}{3}  \log \left[\frac{L }{ \pi } \sin \left(\frac{ \pi  \left(X_1-X_2\right)}{L}\right)\right]   +  \frac{c}{6} \log \left[\frac{\sin \left[\pi \alpha_{\mathcal{O}}\right]}{\pi \alpha_{\mathcal{O}}}\right]\\
       &+\frac{c}{6}\begin{cases}
            \log \left[ \frac{8 \pi t^2}{L  \epsilon  \left(\cot \left(\frac{\pi  X_2}{L}\right)-\cot \left(\frac{\pi  X_1}{L}\right)\right)}\right] & ~\text{for}~ i=1\\
             \log \left[ \frac{L }{2   \pi  \epsilon  \left(\cot \left(\frac{\pi  X_2}{L}\right)-\cot \left(\frac{\pi  X_1}{L}\right)\right)}\right]& ~\text{for}~ i=2\\
       \end{cases}.
\end{split}
\ee
In the system for $i=1$, for the large time region, the entanglement entropy logarithmically increases with $t$ as in \cite{2014arXiv1405.5946C}. 
However, the system considered there is different from the one considered in this paper.
There, the system is an infinite spatial interval, and the subsystem is taken to be half of it.
In that case, since the dimension of Hilbert space associated with the subsystem may be infinite, the entanglement entropy associated with that subsystem can logarithmically increase forever.
In contrast, the system considered in this paper is a finite spatial circle.
Hence, the dimension of the Hilbert space associated with any subsystem may be finite after we introduce a parameter taming the UV divergence. 
Therefore, the time dependence of $S_{A,i=1}$ suggests that the holographic computation may break down for the large time regime.

In the system for $i=2$, the entanglement entropy saturates to a certain value in the large-time regime.
From the time dependence of $S_{A,i=2}$, we can see that the dependence of $S_{A,i=2}$ on the local operator is not suppressed by the logarithmic growth with time.
This suggests that the information about the local operator is not washed out by the time evolution induced by the holographic CFT Hamiltonian on the curved background.

By exploiting the formula in (\ref{eq:entanglement-entropy-and-energy-density}), we will explain the time dependence of $S_{A,i=1,2}$.
Divide $S_{A,i=1,2}$ into $S^{\text{Int.}}_{A,i=1,2}$ and $S^{\text{Loc.}}_{A,i=1,2}$ as in 
\be
S_{A,i=1,2}=S^{\text{Int.}}_{A,i=1,2}+S^{\text{Loc.}}_{A,i=1,2},
\ee
where $S^{\text{Int.}}_{A,i=1,2}$ and $S^{\text{Loc.}}_{A,i=1,2}$ are defined as
\be
\begin{split}
    &S^{\text{Int.}}_{A,i=1,2}=\min\left[\frac{c}{6} \log \left[ 8 \left| \sinh \left[  \f{C}{2} \int_A \sqrt{T(z)} dz \right] \right|\right], \log{\left[8\left| \sinh \left[  \f{C}{2} \int_{\overline{A}} \sqrt{T(z)} dz \right] \right| \right]} \right], \\
     &S^{\text{Loc.}}_{A,\alpha=1,2}=-\frac{c}{6} \log \left[ \delta^2 \left|C\right|^2 \left[T(z_1) \overline{T}(z_1)\right]^{\f{1}{4}} \left[T(z_2) \overline{T}(z_2)\right]^{\f{1}{4}} \right], \\
     &C=-\f{\sqrt{1-24 h_{\mathcal{O}} / c}}{\sqrt{h_{\mathcal{O}}}}.
\end{split}
\ee
Here, $S^{\text{Int.}}_{A,i=1,2}$ is the function determined by the energy-momentum tensors in $A$ or $\overline{A}$, while $S^{\text{Loc.}}_{A,i=1,2}$ is the one determined by the energy-momentum tensors at the boundary points of $A$. To study the time-ordering effect influences the bipartite entanglement, closely look at the behavior of the chiral and anti-chiral energy densities on the plane. 

We begin by taking the small $\epsilon$ expansion on the plan, and then at the leading order, the expectation values of energy-densities are given by
\be
\begin{split}
	&\left\langle T(z_X) \right\rangle_{i=1} \approx 
	\frac{ \pi^2 L^2 h_{\mathcal{O}} \epsilon^2 }
	{ \left(L^2 + 16 \pi^2 t^2\right)^2 \sin^4\left( \dfrac{\pi X}{L} - \phi \right) } \\[6pt]
	&\left\langle \bar{T}(\bar{z}_X) \right\rangle_{i=1} \approx 
	\frac{ \pi^2 L^2 h_{\mathcal{O}} \epsilon^2 }
	{ \left(L^2 + 16 \pi^2 t^2\right)^2 \sin^4\left( \dfrac{\pi X}{L} - \psi \right) } \\[6pt]
	&\left\langle T(z_X) \right\rangle_{i=2} \approx 
	\frac{ \pi^2 L^2 h_{\mathcal{O}} \epsilon^2 }
	{ L^4 \sin^4\left( \dfrac{\pi X}{L} - \phi \right) } \\[6pt]
	&\left\langle \bar{T}(\bar{z}_X) \right\rangle_{i=2} \approx 
	\frac{ \pi^2 L^2 h_{\mathcal{O}} \epsilon^2 }
	{ L^4 \sin^4\left( \dfrac{\pi X}{L} - \psi \right) }.
\end{split}
\ee
Then, by taking $t$ to be large, we can approximate those expectation values as 
\be\label{eq:Energy-density-large-time}
\begin{split}
	\left\langle T(z_X) \right\rangle_{i=1} 
	&\approx \frac{L^2 h_{\mathcal{O}} \epsilon^2}
	{256\,\pi^2\,t^4\,\sin^4\left( \dfrac{\pi X}{L} \right)} \\[4pt]
	\left\langle \bar{T}(\bar{z}_X) \right\rangle_{i=1} 
	&\approx \frac{L^2 h_{\mathcal{O}} \epsilon^2}
	{256\,\pi^2\,t^4\,\cos^4\left( \dfrac{\pi X}{L} \right)} \\[4pt]
	\left\langle T(z_X) \right\rangle_{i=2} 
	&\approx \frac{\pi^2 h_{\mathcal{O}} \epsilon^2}
	{L^2 \sin^4\left( \dfrac{\pi X}{L} \right)} \\[4pt]
	\left\langle \bar{T}(\bar{z}_X) \right\rangle_{i=2} 
	&\approx \frac{\pi^2 h_{\mathcal{O}} \epsilon^2}
	{L^2 \cos^4\left( \dfrac{\pi X}{L} \right)}.
\end{split}
\ee

Thus, in this limit, for $i=1$, the chiral and anti-chiral energy densities far from the energy peak approximately decay  as $\f{1}{t^4}$, while those for $i=2$ are approximately given by the constant. This suggests that the excitation for $i=1$ is entirely distributed in the system because the energy densities, far from the energy peak, decay according to the propagation of the energy peak.
In contrast, the excitation for $i=2$ is localized because the energy distribution, far from the energy peak, is independent of time.  
For $i=1$, due to the decay in time of those energy densities, $S_{A,i=1}$ logarithmically grows {with} $t$, while for $i=2$, $S_{A,i=2}$ becomes constant since both of $S^{\text{Int.}}_{A,\alpha=1,2}$ and $S^{\text{Loc.}}_{A,\alpha=1,2}$ become independent of $t$. 
From these differences between $i=1$ and $i=2$, we can see that the time evolution for $i=1$ induces the widely distributed excitation, and this excitation induces the logarithmic growth of the entanglement entropy, while that for $i=2$ induces the local excitation, which causes the entanglement entropy to eventually saturate at a finite value.

\subsubsection{case b}
\begin{figure}[htbp!]
	\centering
	\includegraphics[width=0.6\textwidth]{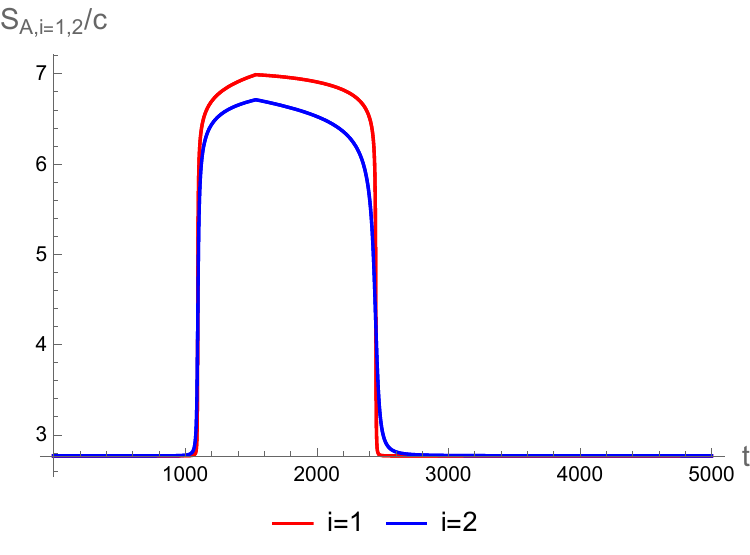} %
	\caption{
		Entanglement entropy as a function of time, with parameters 
		$L = 10000$, $c = 1000$, $h_{\mathcal{O}} = \frac{c}{24} + 10$, 
		$\delta = 1$, $x = 2500$, $X_1 = 4500$, $X_2 = 1000$, 
		and $\epsilon = 10$.	}
	\label{fig:case_b_ee}
\end{figure}

 As shown in Fig.~\ref{fig:case_b_ee}, we now analyze the time dependence of $S_{A,i=1,2}$ for case b, where the behavior is simpler than in case a.
Before $t= t_{X_1}= \frac{L \cot \left(\frac{2 \pi  X_1}{L}\right)}{4 \pi }$, $A$ contains the quasiparticles induced by the local operator.
In the time interval, $t_{X_1}<t<t_{X_2}$, only one of the quasiparticles is in $A$, so that the entanglement between them can contribute to $S_A$.
In the late time region, $t>t_{X_2}$, both quasiparticles are out of $A$. 
Following this physical picture, we obtain the time dependence of $S_{A,i}$ as
\be
\begin{split}
       &S_{A,i=1,2}\approx \frac{c}{3}  \log \left[\frac{L }{ \pi } \sin \left(\frac{ \pi  \left(X_1-X_2\right)}{L}\right)\right]   
 +  \begin{cases}
0 &  t_{X_1}>t>0\\
\frac{c}{6} \log \left[\frac{\sin \left[\pi \alpha_{\mathcal{O}}\right]}{\pi \alpha_{\mathcal{O}}}\right]-\frac{c}{6}  \log \left[\epsilon g_{i=1,2}\left( t\right)\right] &   t_{-}>t>t_{X_1} \\
\frac{c}{6} \log \left[\frac{\sin \left[\pi \alpha_{\mathcal{O}}\right]}{\pi \alpha_{\mathcal{O}}}\right]-\frac{c}{6}  \log \left[-\epsilon f_{i=1,2}\left( t\right)\right] &   t_{X_2}>t>t_{-} \\
0 & t>t_{X_2}\\
    \end{cases},\\
\end{split}
\ee
where
$t_{-}=\frac{L \sqrt{\tan \left(\frac{2 \pi  \left(X_1-\frac{L}{4}\right)}{L}\right) \tan \left(\frac{2 \pi  \left(\frac{L}{4}-X_2\right)}{L}\right)}}{4 \pi }$. 
The exchange time, $t_-$, is defined as the time for the analytically-continued entanglement entropy with $g_i(t)$ to exchange the dominance with that with $f_i(t)$. 
\subsection{Process of the destroying quasiparticles \label{sec:for-general-cr}}
Move on to the second main research topic of this paper.
First, we clarify the definition of quasiparticles more.
In this paper, the quasiparticles are defined as the entangled excitations.
When the local operator is inserted into the system, two quasiparticles emerge at the spatial point where the local operator is inserted, and they form an entangled pair.
During the time evolution, the peaks of chiral and anti-chiral energy densities indicate the locations of these quasiparticles.
Assume that we insert the local operator into $x=\f{L}{4}$ of the vacuum state, and then evolve the system with $H_{2-\text{SSD}}$.
Then, at $t=\left|\f{L \cot{\left(\f{2\pi x}{L}\right)}}{4\pi}\right|$, the sharp energy-density peak exists at $x$, where we assume that $0\le x \le \f{L}{2}$.
This indicates that at this time, the location of quasiparticles is at $x$.
The growth of entanglement entropy reported above can be interpreted as the contribution from the entanglement between this entangled pair.
If only one of the quasiparticles is in $A$, the entanglement between those two quasiparticles contributes to the entanglement entropy for $A$.
In two-dimensional holographic CFTs, although this effective description cannot describe how much the entanglement entropy increases, it can qualitatively describe the behavior of the entanglement entropy.
To further investigate if the increase in the entanglement entropy is due to the non-local object, i.e. the entangled pair, we consider the following sequence of processes:
\begin{itemize}
    \item In the first process, we evolve the system with $H_{2-\text{SSD}}$. 
    During the time evolution induced by $H_{2-\text{SSD}}$, one of the quasiparticles generated by the local operator propagates to $x=0$, while the other to $x=\f{L}{2}$. Therefore, for the large time region, the quasiparticles should accumulate around $x=0$ or  $x=\f{L}{2}$.
    In other words, in this time region, the energy density has sharp peaks around $x=0$ and $\f{L}{2}$.
    \item In the second process, we turn on the damping factor, $e^{-\epsilon_2 H_{\text{SSD}}}$. Define the energy density of $H_{\text{SSD}}$ as 
    \be
    H_{\text{SSD}}=\int^{L}_0 dx \mathcal{E}(x), \text{where}~\mathcal{E}(x)=2\sin^2{\left(\f{\pi x}{L}\right)}e(x).
    \ee
    Since around $x=0$, the energy density is approximately zero, $\mathcal{E}(x\approx 0)\approx 0$, this damping factor weakly affects the spatial region around $x=0$.
    In contrast, this damping factor strongly affects the spatial region around $x=\f{L}{2}$ because $\mathcal{E}(x\approx \f{L}{2})\approx 2 e(x)$.
    If the value of $\epsilon_2$ becomes large, the energy density around $x=\f{L}{2}$ will be suppressed more than that around $x=0$.
    Hence, this process may destroy the quasiparticle around $x=\f{L}{2}$, while keeping the one around $x=0$. 
\end{itemize}

In this section we investigate how the damping factor influences the late-time behaviour of the entanglement entropy in case~a.  
At late times, one quasiparticle sits near \(x=0\), whereas its partner is located around \(x=\tfrac{L}{2}\).  
Since the subsystem of interest is the spatial region surrounding \(x=0\), the quasiparticle at \(x=0\) remains inside the subsystem \(A\), while the one near \(x=\tfrac{L}{2}\) is effectively removed by the damping factor.  
In other words, a quasiparticle that is far from the subsystem under consideration may be destroyed.
To quantify the resulting decrease in entanglement entropy due to the decay of quasiparticle, we introduce the local internal energies centred at  
\begin{equation}
  x_L=\frac{L}{2\pi}\cot^{-1}\!\bigl(4\pi\,t/L\bigr), 
  \qquad
  x_R=\frac{L}{2}-\frac{L}{2\pi}\cot^{-1}\!\bigl(4\pi\,t/L\bigr).
\end{equation}
Let us define those internal energies as
\begin{equation}
\begin{split}
  E_{R;i=3,4} &= \int_{x_R-l}^{\,x_R+l} dx\,\bigl[T(z)+\bar{T}(\bar z)\bigr],\\
  E_{L;i=3,4} &= \int_{x_L-l}^{\,x_L+l} dx\,\bigl[T(z)+\bar{T}(\bar z)\bigr],
\end{split}
\end{equation}
where \(l\) is a small positive parameter, and chosen here to be \(l = L/10^6\).

\begin{figure}[htbp]
\centering
{\includegraphics[width=0.8\columnwidth]{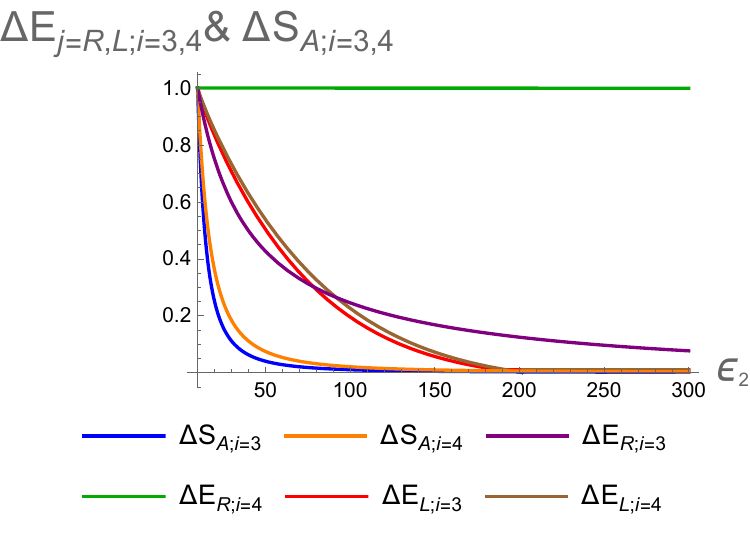}}
	\caption{ 
     The decay rates of the internal energy and entanglement entropy as the function of $\epsilon_2$. Blue and orange solid lines illustrate the $\epsilon_2$ dependence of $\Delta E_{R; i=3,4}$, while purple and green lines illustrate that of $\Delta E_{L; i=3,4}$. The red and brown solid lines illustrate the $\epsilon_2$ dependence of $\Delta S_{A; i=3,4}$. Here, the parameters are set as follows: $t = 100000$, $c = 1000$, $h_{\mathcal{O}} = 1000$, $\epsilon_2^0=5$, $\epsilon_1 = 10$, $L = 10000$, $x = 2500$,$x_{L}=12.6649$, $x_R=4987.34$, $X_1 = 8750$, and $X_2 = 1250$. 
   }\label{Fig:Energy_density_Ratio}
\end{figure}

Then, to investigate how much the damping factor suppresses the internal energy and entanglement entropy, we define those late‑time decay rates as 
\be
	\Delta S_{A; i=3,4} = 
	\frac{S_{A; i=3,4} - S_{A; \mathrm{vac}}}{
		\left(S_{A; i=3,4} - S_{A; \mathrm{vac}}\right)\big|_{\epsilon_2 = \epsilon_2^0}}, \quad
	\Delta E_{j; i=3,4} = 
	\frac{E_{j; i=3,4}}{E_{j; i=3,4}\big|_{\epsilon_2 = \epsilon_2^0}} 
	\quad (j = L, R),
\ee

where $S_{A; vac}$ denotes the vacuum entanglement entropy for the subregion $A$, and $\epsilon_2^0$ is a positive constant \footnote{Ideally, $\epsilon^0_2$ should be set to be zero and we should compare $\epsilon_2$ dependence of $\Delta S_{A;i}$ with that of $\Delta E_{j;i}$. However, we faced some difficulties when we numerically calculate $\Delta E_{j;i}$ for small $\epsilon^0_2$.
Therefore, we introduced $\epsilon_2^0$ for Figs. \ref{Fig:Energy_density_Ratio} and  \ref{Fig:Mutual34}.
However, we claim that the $\epsilon_2$ dependence of those quantities should not strongly depend on $\epsilon^0_2$. }.

In Fig. \ref{Fig:Energy_density_Ratio}, we show $\Delta S_{A; i}$ and $\Delta E_{j; i}$ for the late-time regime as the functions of $\epsilon_2$.
In this figure, we set $l=\f{L}{10^6}$.
From this figure, we can see that those decay rates monotonically decrease in $\epsilon_2$. 
To closely investigate the $\epsilon_2$-dependence of $\Delta E_{j; i}$, let us compare $\Delta E_{j=L; i}$ to $\Delta E_{j=R; i}$.
In both $i=3$ and $4$, $\Delta E_{j=R; i}$ decreases more rapidly than $\Delta E_{j=L; i}$ according to the increase in $\epsilon_2$. 
Consequently, with the large $\epsilon_2$,  $\Delta E_{j=L; i}$ approximately vanishes, while  $\Delta E_{j=R; i}$ does not.
This suggests that the damping factor, $e^{-\epsilon H_{\text{SSD}}}$, suppresses the quasiparticle around $x=\f{L}{2}$ more drastically than around $x=0$. 
Let us compare the $\epsilon_2$ dependence of $\Delta E_{j; i=3}$ with that of $\Delta E_{j; i=4}$.
The behavior of $\Delta E_{R; i=4}$ significantly differs from
that of $\Delta E_{R; i=3}$.
According to the growth of $\epsilon_2$, $\Delta E_{R; i=3}$ monotonically decreases, while  $\Delta E_{R; i=4}$ is approximately one.
One possible interpretation is that since for $i=3$, excitation is delocalized in the entire system, while for $i=3$ is well-localized, the damping factor influences $\Delta E_{R; i=3}$ more strongly than $\Delta E_{R; i=4}$, so that $\Delta E_{R; i=3}$ decays faster than  $\Delta E_{R; i=4}$.

Move on to closely looking at the $\epsilon_2$ dependence of $\Delta S_{A;i}$.
From the $\epsilon_2$ dependence of $\Delta S_{A;j}$ in Fig. \ref{Fig:Energy_density_Ratio}, we can observe that $\Delta S_{A;j}$ monotonically decreases with $\epsilon_2$. 

Then, compare the $\epsilon_2$ dependence of $\Delta S_{A;i}$ with that of $\Delta E_{j;i}$.
According to the growth of $\epsilon_2$, $\Delta E_{L;i}$ and $\Delta S_{A;i}$ vanish, while $\Delta E_{R;i}$ remains finite.
This suggests that the damping factor, $e^{-\epsilon H_{\text{SSD}}}$, may suppress the quasiparticle far from $A$, while keeping quasiparticle in $A$.
Then, the decay of the quasiparticle far from $A$ may cause the decay of the entanglement entropy for $A$.
As a consequence, the increase in entanglement entropy from the vacuum one may be due to the entanglement between quasiparticles around $x=0$ and $x=\f{L}{2}$.
\section{Mutual information \label{sec:mutual-information}}
In the last section, by exploiting entanglement entropy, we investigated how the local operator induces the bipartite entanglement, gave a physical picture qualitatively describing the time dependence of the entanglement entropy, explored how the time ordering effect on the bipartite entanglement, and then investigated the relation between the bipartite entanglement and the energy density and the suppressing effect of the damping factor on the energy densities and entanglement entropy.
In this section, we will investigate how the local operator influences the non-local correlation of the system, whether the quasiparticle picture proposed can describe the behavior of the non-local correlation induced by the local operator, how the time ordering influences the non-local correlation, and then how the damping factor affects this correlation, by exploring the time dependence of the mutual information defined as in (\ref{eq:def-of-MI}).

Suppose that two subsystems for the mutual information, $A$ and $B$, are the distant intervals defined as
\be
\begin{split}
    A=[X_2,X_1], ~ B=[Y_1,L]\cup [0,Y_2],
\end{split}
\ee
where we assume that $Y_2<X_2<L/2<X_1<Y_1$.
For simplicity, we assume that the location of the local operator is $x=\f{L}{4},$ and the spatial interval between $x=Y_2$ and $x=\f{L}{4}$ is larger than the one between $x=X_2$ and $x=\f{L}{4}$, $X_2-\f{L}{4}< \f{L}{4}-Y_2$.
The subsystem, $A$, contains $x=\f{L}{2}$, the one of the fixed points for $H_{2-\text{SSD}}$, while $B$ contains $x=0$, the other one.

We begin by investigating the non-local correlation induced by the insertion of the local operator into the vacuum state by exploiting the mutual information between $A$ and $B$. 
To this end, we consider the systems for $i=1$ and $2$.
Suppose that the time dependence of $I_{A,B}$ for $i=1$ and $2$ qualitatively follows the quasiparticle picture.
In the quasiparticle picture for the time evolution induced by $H_{2-\text{SSD}}$, one of the quasiparticles induced by the local operator propagates to $x=0$, while the other to $x=\f{L}{2}$.
Consequently, for the large time region, $A$ contains the single quasiparticle, while $B$ contains the other.
Therefore, the increase due to the quasiparticle in $S_{A\cup B}$ is zero because the entangled pair is contained by $A\cup B$.
In contrast, since $A$ and $B$ contain a single quasiparticle, $S_A$ and $S_B$ increase thanks to the entanglement between them.

\begin{figure}[htbp]          
	\centering
	\includegraphics[width=0.8\linewidth]{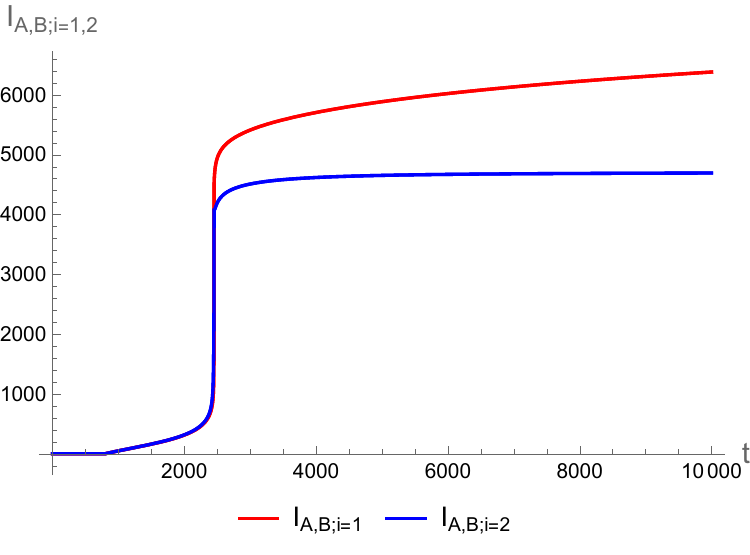}
	\caption{The behaviors of the mutual information of $i=1,2$ as the function of $t$. Here, we set $L=10000$, $\epsilon_1=10$, $Y_2=500, X_2=3000, X_1=5500, Y_1=8000$ and $h_{\mathcal{O}}=c=1000$.}
	\label{Fig:Mutual-Information}
\end{figure}

Then, we move on to the analysis of $I_{A,B}$ for the systems, $i=1$ and $2$. 
Since we consider the spatially-periodic system in pure state, $S_{A\cup B}=S_{\overline{A\cup B}}$, where $\overline{A\cup B}$, the complement to $A\cup B$, is defined as
\be
\overline{A\cup B} =C\cup D,~C=[Y_2, X_2],~D= [X_1,Y_1].
\ee
Since in the early time region, $-\f{L \cot{\left(\f{2\pi X_2}{L}\right)}}{4\pi}>t>0$, $\overline{A\cup B}$ contains both quasiparticles, the entanglement between them does not contribute to $S_{A\cup B}$. 
Hence, $S_{A\cup B}$ may result in the vacuum entanglement entropy, i.e., the time-independent entanglement entropy.
\if[0]
\be \label{eq:initial-SAB}
\begin{split}
   \textcolor{red}{ S_{A\cup B}=\begin{cases}
       \sum_{\alpha =A,B} \f{c}{3}\log{\left[\f{L}{\pi}\sin{\left[\f{\pi l_{\alpha}}{L}\right]}\right]}~&~\text{for}~ \f{\prod_{\alpha=A,B}\sin{\left[\f{\pi l_{\alpha}}{L}\right]}}{\prod_{\beta=C,D}\sin{\left[\f{\pi l_{\beta}}{L}\right]}}>1\\
       \sum_{\alpha =C,D} \f{c}{3}\log{\left[\f{L}{\pi}\sin{\left[\f{\pi l_{\alpha}}{L}\right]}\right]}~&~\text{for}~ \f{\prod_{\alpha=A,B}\sin{\left[\f{\pi l_{\alpha}}{L}\right]}}{\prod_{\beta=C,D}\sin{\left[\f{\pi l_{\beta}}{L}\right]}}<1\\
    \end{cases},}
\end{split}
\ee
where $l_{\alpha}$ is defined as
\be
l_{A}=X_1-X_2, l_{B}=L-(Y_1-Y_2), l_{C}=X_2-Y_2, l_{D}=Y_1-X_1.
\ee
\fi
In the time interval, $\f{L \cot{\left(\f{2\pi Y_2}{L}\right)}}{4\pi}>t>-\f{L \cot{\left(\f{2\pi X_2}{L}\right)}}{4\pi}$, the entanglement between quasiparticles contributes to $S_{A\cup B}$.
In the late time region, $t>\f{L \cot{\left(\f{2\pi Y_2}{L}\right)}}{4\pi}$, $S_{A\cup B}$  may become the time-independent one since the entanglement between the quasiparticles does not contribute.
The time dependence of $S_A$ and $S_B$ is similar to the one reported in Section 
 \ref{sec:Warm-up}.
By combining the time-dependence of $S_A$ and $S_B$ with $S_{A\cup B}$, we obtain the time dependence of $I_{A,B}$, and show it as a function of $t$ in Fig. \ref{Fig:Mutual-Information}.
This figure shows that $I_{A,B}$ is zero before $t=\f{L \cot{\left(\f{2\pi X_2}{L}\right)}}{4\pi}$, while after $t=\f{L \cot{\left(\f{2\pi X_2}{L}\right)}}{4\pi}$, $I_{A,B;i=1,2}$ drastically grows in $t$.
For large time regime, $I_{A,B;i=1}$ logarithmically grows in $t$, while  $I_{A,B;i=2}$ saturates to constant.
At the leading order in the large time expansion, the late-time behavior of $I_{A,B;i=1,2}$ is approximated by
\be
   \begin{split}
  I_{A,B;i=1,2}&\approx  \frac{c}{3} \log \left[\frac{\sin \left[\pi \alpha_{\mathcal{O}}\right]}{\pi \alpha_{\mathcal{O}}}\right]\\
   	&+\frac{c}{6}\begin{cases}
   		\log \left[ \frac{8 \pi t^2}{L  \epsilon  \left(\cot \left(\frac{\pi  X_2}{L}\right)-\cot \left(\frac{\pi  X_1}{L}\right)\right)}\right] +\log \left[ \frac{8 \pi t^2}{L  \epsilon  \left(\cot \left(\frac{\pi  Y_2}{L}\right)-\cot \left(\frac{\pi  Y_1}{L}\right)\right)}\right]& ~\text{for}~ i=1\\
   		\log \left[ \frac{L }{2   \pi  \epsilon  \left(\cot \left(\frac{\pi  X_2}{L}\right)-\cot \left(\frac{\pi  X_1}{L}\right)\right)}\right]+\log \left[ \frac{L }{2   \pi  \epsilon  \left(\cot \left(\frac{\pi  Y_2}{L}\right)-\cot \left(\frac{\pi  Y_1}{L}\right)\right)}\right]& ~\text{for}~ i=2\\   	\end{cases},
   \end{split}
\ee
where we reported on the pieces at $\log{\left(1/\epsilon\right)}$ and those depending on the $\alpha_{\mathcal{O}}$.
The formula in (\ref{eq:entanglement-entropy-and-energy-density}) states that the logarithmic growth of $I_{A,B;i=1}$ is induced by the widely distributed excitation, while the local excitation for $i=2$ induces the constant late-time value of $I_{A,B;i=2}$. 

\begin{figure}[htbp]          
	\centering
	\includegraphics[width=0.8\linewidth]{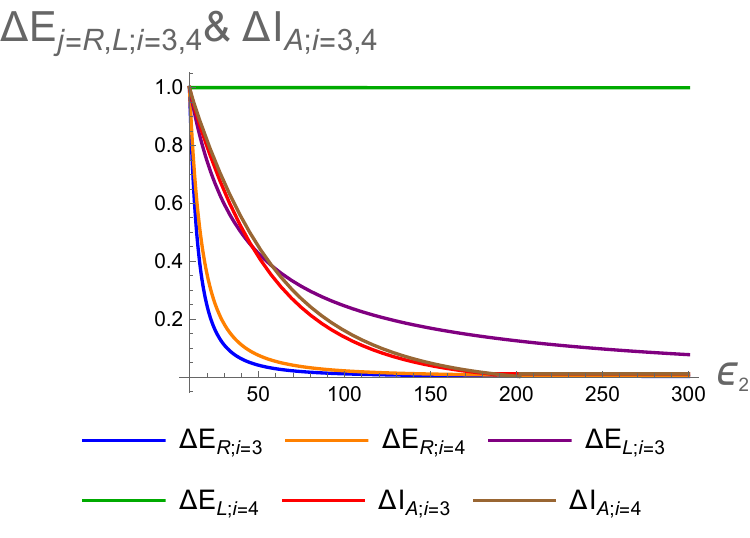}
	\caption{The $\epsilon_2$-dependence of $\Delta E_{j=R,L;i=3,4}$ and $\Delta S_{i=3,4}$ . Here, we take $t = 100000$, $L=10000$, $\epsilon_2^0=5$, $\epsilon_1=10$, $Y_2=1250, X_2=3750, X_1=6250, Y_1=8750, x_{L}=12.6649, x_R=4987.34$ and $h_{\mathcal{O}}=c=1000$.}
	\label{Fig:Mutual34}
\end{figure}

Then, we will investigate the effect of the damping factor, $e^{-\epsilon_2 H{\text{SSD}}}$, on the non-local correlation by exploiting the mutual information between $A$ and $B$.
In the quasiparticle picture, for the large time regime, one quasiparticle is around $x=0$, while the other is around $x=\f{L}{2}$.
To investigate how the damping factor suppresses the non-local correlation, define the decay rate of $I_{A,B}$ as 
\be
\Delta I_{A,B;i=3,4}=\f{ I_{A,B;i=3,4}}{ I_{A,B;i=3,4}|_{\epsilon_2=\epsilon_2^0}}
\ee
again, here $\epsilon_2^0$ is a positive constant.
In Fig. \ref{Fig:Mutual34}, 
we show $\Delta I_{A,B;i=3,4}$ and $\Delta E_{j;i=3,4}$ as the function of $\epsilon_2$.
From this figure, we can see that $I_{A,B}$ monotonically decrease with $\epsilon_2$, and then for large-$\epsilon_2$ regime, $\Delta I_{A,B}$ and $\Delta E_{R;i}$  vanishes, while $\Delta E_{L;i}$ remains finite.
This suggests that the non-local correlation between $A$ and $B$ may decay due to the decay of the quasiparticle around $x=0$. Especially, the $\epsilon_2$ dependence of $\Delta S_{A;i=4}$, $I_{A,B;i=4}$, and $\Delta E_{j=L,R;i=4}$ in Figs. \ref{Fig:Energy_density_Ratio} and \ref{Fig:Mutual34} more clearly suggests that the decay of bi-partite entanglement and non-local correlation can be induced by that of the quasiparticle around $x=0$. The quantities for $i=3$ decay more drastically than those for $i=4$.
One possible interpretation of this difference between them is that since the excitation for $i=3$ is distributed more widely than that for $i=4$, the damping factor suppresses the excitation for $i=3$ more drastically than that for $i=4$.

We will close this section with an interpretation on $S_{A\cup B}$ by the usage of (\ref{eq:entanglement-entropy-and-energy-density}).
In $2$d holographic CFTs, the time dependence of $S_{A\cup B}$ is determined by the minimal geodesic length. 
In this case, we have two geodesics, and the minimal one is determined by
\be
\begin{split}
    \mathcal{R}=\f{\left|\sinh{\left[\f{C}{2}\int_{A} \sqrt{T(z)}dz\right]}\right|\left|\sinh{\left[\f{C}{2}\int_{B} \sqrt{T(z)}z\right]}\right|}{\left|\sinh{\left[\f{C}{2}\int_{C} \sqrt{T(z)}dz\right]}\right|\left|\sinh{\left[\f{C}{2}\int_{D}\sqrt{T(z)}\right]}\right|},
\end{split}
\ee
where for $\mathcal{R}<1$, the length of geodesic associated with $\int_{A} \sqrt{T(z)}dz$ $\int_{B} \sqrt{T(z)}dz$, and anti-holomorphic parts of them determines the behavior of $S_{A\cup B}$, while for $\mathcal{R}>1$, the one associated with $\int_{C} \sqrt{T(z)}dz$, $\int_{D} \sqrt{T(z)}dz$, and anti-holomorphic parts of them does.
The integrals \( \int \sqrt{T(z)}\,dz ~\left( \int \sqrt{\overline{T}(\overline{z})}\,d\overline{z}\right) \) are taken along  the contour \( z = e^{\frac{2\pi iX}{L}}~(\overline{z} = e^{\frac{-2\pi iX}{L}}) \).  
Then, the sine-hyperbolic function reduces to 
\be
\begin{split}
    &\left|\sinh{\left[\f{C}{2}\int_{\mathcal{V}=A,B,C,D} dz \sqrt{T(z)} \right]}\right|\\
    &=
    \begin{cases}
        \sin{\left(\f{C}{2}\int_{\mathcal{V}} dX \sqrt{T(w=iX)}\right)}\sin{\left(\f{C}{2}\int_{\mathcal{V}} dX \sqrt{\overline{T}(\overline{w}=-iX)}\right)}~&~\text{For}~\f{c}{24}>h_{\mathcal{O}}\\
        \sinh{\left(\f{D}{2}\int_{\mathcal{V}} dX \sqrt{T(w=iX)}\right)}\sinh{\left(\f{D}{2}\int_{\mathcal{V}} dX \sqrt{\overline{T}(\overline{w}=-iX)}\right)}~&~\text{For}~h_{\mathcal{O}}>\f{c}{24}\\
    \end{cases},
\end{split}
\ee
where $D=\sqrt{\f{24 h_{\mathcal{O}}}{c}-1}$.
Thus, the behavior of $\mathcal{R}$ is determined by the internal-energy-like quantities associated with the subregions,  $\int_{\mathcal{V}} dX \sqrt{T(iX)}$ and $\int_{\mathcal{V}} dX \sqrt{\overline{T}(-iX)}$

\section{Gravitational Dual Geometry}\label{sec:gravity-dual}

In this section, we will investigate the gravitational dual geometry corresponding to the systems discussed in the previous sections. 
As shown in \eqref{eq:metric-banados}, the geometry is the function determined by the chiral and anti-chiral energy-momentum tensors.
 As pointed out in Refs.~\cite{Abajian:2023jye, Abajian:2023bqv}, the coordinates for the Bañados geometry can not cover the deep region of the gravitational dual. 
 For example, in the case of the BTZ black hole, those coordinates cannot reach the horizon of the black hole.  
Here, we consider $h_{\mathcal{O}}>c/24$, where the gravitational dual is approximately given by the black hole geometry.
To investigate the dynamical properties of the gravitational dual, we will closely explore the behavior of the black hole horizon.
To this end, we will change the coordinates to the one covering the entire region, i.e. the global coordinate, where the geometry is the BTZ black hole, and then, we will investigate the spacetime dependence of the BTZ black hole.  We refer to Ref.~\cite{Das:2024lra} as one of the prior works where a similar method of horizon identification is discussed.

\subsection{BTZ Geometry}

Let us begin by reviewing  Euclidean  BTZ black hole with mass $M=r_{+}^2+r_{-}^2$ and angular momentum $J=2 r_{+} r_{-}$, where $r_+$ and $r_-$ denote the outer and inner horizons and $r_-<r_+$.
This black hole geometry in the global coordinate is presented as 
\begin{equation} \label{eq:BTZ-black-hole-for-general}
	d s_E^2=\frac{\left(r^2-r_{+}^2\right)\left(r^2-r_{-}^2\right)}{r^2 \ell^2} d \xi^2+\frac{r^2 \ell^2}{\left(r^2-r_{+}^2\right)\left(r^2-r_{-}^2\right)} d r^2+r^2\left(d \phi+\frac{r_{+}\left(i r_{-}\right)}{r^2} d \xi\right)^2,
\end{equation}
where $\phi \sim \phi+2 \pi$.
The map from $(v^{+}, v^{-}, u)=(v,  \overline{v}, u)$ to $(r, \phi, \xi)$ is given by  
\begin{equation}\label{TS:AdS-BTZ}
	v^{ \pm}=\left(\frac{r^2-r_{+}^2}{r^2-r_{-}^2}\right)^{\frac{1}{2}} e^{2 \pi T_{ \pm}(\phi \pm i\xi)}, \quad u=\left(\frac{r_{+}^2-r_{-}^2}{r^2-r_{-}^2}\right)^{\frac{1}{2}} e^{\pi T_{+}(\phi+i\xi)+\pi T_{-}(\phi-i\xi)},
\end{equation}
where we defined left and right temperatures as 
\begin{equation}
	T_L=T_{+}=\frac{1}{2 \pi}\left(r_{+}+r_{-}\right), \quad T_R=T_{-}=\frac{1}{2 \pi}\left(r_{+}-r_{-}\right).
\end{equation}
In terms of $h_{\mathcal{O}}$ and $c$, those temperatures are expressed as
\begin{equation}
	T_L=\frac{1}{2 \pi} \sqrt{\frac{24 h_{\mathcal{O}}}{c}-1}, \quad T_R=\frac{1}{2 \pi} \sqrt{\frac{24 \overline{h}_{\mathcal{O}}}{c}-1}.
\end{equation}
The coordinate transformation in (\ref{TS:AdS-BTZ}) maps from BTZ metric to locally $\mathrm{AdS}_3$,
\begin{equation}
	d s^2=\frac{d u^2+d v d \overline{v}}{ u^2}.
\end{equation}

Let us consider the gravity duals of the systems considered in the previous sections.
As in the previous sections, take $h_{\mathcal{O}}=\overline{h}_{\mathcal{O}}$.
Then, there is only one outer horizon at  $r_{+}={r_{h_\mathcal{O}}}$ , and inner horizon has shrunk to $ r_-=0$. 
Consequently, the BTZ geometry in (\ref{eq:BTZ-black-hole-for-general}) is simplified as 
\begin{equation} \label{eq:BTZ-simplified}
	d s_E^2= \ell^2 \left( \frac{\left(
		r^2-{r_{h_\mathcal{O}}}^2\right)}{ \ell^2} \frac{d \xi^2}{\ell^2}+\frac{ \ell^2}{\left(r^2-{r_{h_\mathcal{O}}}^2\right)}  \f{d r^2}{\ell^2}+\frac{\ell^2}{r^2} {d \phi}^2\right).
\end{equation}
Then, we map from the Bañados and BTZ geometries to the Poincaré AdS$_3$.  
The map from the BTZ geometry to Poincaré AdS$_3$ is given as
\begin{equation}\label{eq:map-btz-ads}
	v = \sqrt{1 - \left( \frac{{r_{h_\mathcal{O}}}}{r} \right)^2} \, e^{\frac{{r_{h_\mathcal{O}}}}{\ell}(\phi + \f{\xi}{\ell})}, \quad
	\overline{v} = \sqrt{1 - \left( \frac{{r_{h_\mathcal{O}}}}{r} \right)^2} \, e^{\frac{{r_{h_\mathcal{O}}}}{\ell}(\phi - \frac{\xi}{\ell})}, \quad
	u = \frac{{r_{h_\mathcal{O}}}}{r} e^{\frac{{r_{h_\mathcal{O}}}}{\ell} \phi}.
\end{equation}
 On the other hand, the functions, \( f(z) \) and \( \overline{f}(\overline{z}) \), map from the Bañados geometry to AdS$_3$.  They are
\begin{equation}\label{eq:banados-ads}
\begin{aligned}
f(z) &= \left( \frac{z - z_{\epsilon}^{\text{new},i}}{\,z - z_{-\epsilon}^{\text{new},i}\,} \right)^{\,i \frac{r_{h_{\mathcal{O}}}}{\ell}}, \quad
\bar{f}(\bar{z}) &= \left( \frac{\bar{z} - \bar{z}_{\epsilon}^{\text{new},i}}{\,\bar{z} - \bar{z}_{-\epsilon}^{\text{new},i}\,} \right)^{-\,i \frac{r_{h_{\mathcal{O}}}}{\ell}}.
\end{aligned}
\end{equation}
 Here, \( f(z) \) and \( \overline{f}(\overline{z}) \) are derived by solving \eqref{eq:ODE-f-T}.

We first map the Ba\~nados geometry to the Poincar\'e patch of AdS$_3$ using~\eqref{eq:banados-ads}, and map the BTZ geometry to the same Poincar\'e patch using~\eqref{eq:map-btz-ads}. Identifying the images and eliminating the intermediate Poincar\'e coordinates yields a direct transformation between the Ba\~nados and BTZ charts. Explicitly, we obtain
\begin{equation}
\begin{aligned}
&\frac{\left( \frac{z - z_{\epsilon}^{\text{new},i}}{\,z - z_{-\epsilon}^{\text{new},i}\,} \right)^{\,i \frac{r_{h_{\mathcal{O}}}}{\ell}} \, \left( \frac{\bar{z} - \bar{z}_{\epsilon}^{\text{new},i}}{\,\bar{z} - \bar{z}_{-\epsilon}^{\text{new},i}\,} \right)^{-\,i \frac{r_{h_{\mathcal{O}}}}{\ell}} \left[-(-i + \f{{r_{h_\mathcal{O}}}}{\ell})^2 y^2 + 4 \left( \frac{z - z_{\epsilon}^{\text{new},i}}{\,z - z_{-\epsilon}^{\text{new},i}\,} \right) \left( \frac{\bar{z} - \bar{z}_{\epsilon}^{\text{new},i}}{\,\bar{z} - \bar{z}_{-\epsilon}^{\text{new},i}\,} \right) \right]}{- (i + \f{{r_{h_\mathcal{O}}}}{\ell})^2 y^2 + 4 \left( \frac{z - z_{\epsilon}^{\text{new},i}}{\,z - z_{-\epsilon}^{\text{new},i}\,} \right) \left( \frac{\bar{z} - \bar{z}_{\epsilon}^{\text{new},i}}{\,\bar{z} - \bar{z}_{-\epsilon}^{\text{new},i}\,} \right)} = e^{2 i \f{{r_{h_\mathcal{O}}}}{\ell} \f{\xi}{\ell}}, \\
&\left( \frac{z - z_{\epsilon}^{\text{new},i}}{\,z - z_{-\epsilon}^{\text{new},i}\,} \right)^{\,i \frac{r_{h_{\mathcal{O}}}}{\ell}} \, \left( \frac{\bar{z} - \bar{z}_{\epsilon}^{\text{new},i}}{\,\bar{z} - \bar{z}_{-\epsilon}^{\text{new},i}\,} \right)^{-\,i \frac{r_{h_{\mathcal{O}}}}{\ell}} = e^{2 \f{{r_{h_\mathcal{O}}}}{\ell} \phi}, \\
&\frac{\left[(1 + \f{{r_{h_\mathcal{O}}}}{\ell}^2) y^2 + 4 \left( \frac{z - z_{\epsilon}^{\text{new},i}}{\,z - z_{-\epsilon}^{\text{new},i}\,} \right)\left( \frac{\bar{z} - \bar{z}_{\epsilon}^{\text{new},i}}{\,\bar{z} - \bar{z}_{-\epsilon}^{\text{new},i}\,} \right)\right]^2}{16 y^2 \left( \frac{z - z_{\epsilon}^{\text{new},i}}{\,z - z_{-\epsilon}^{\text{new},i}\,} \right) \left( \frac{\bar{z} - \bar{z}_{\epsilon}^{\text{new},i}}{\,\bar{z} - \bar{z}_{-\epsilon}^{\text{new},i}\,} \right)} = \f{r^2}{\ell^2}.
\end{aligned}
\end{equation}

 These equations define a one-to-one mapping from the Bañados geometry with coordinates $(z, \bar{z}, y)$ to $(\xi, \phi, r)$ of the BTZ black hole. In particular, the third equation encodes the radial structure of the geometry, and the horizon is located at $r={r_{h_\mathcal{O}}}$. 
To match the systems under consideration, the asymptotic geometry near the boundary should be given by the  Poincaré AdS$_3$.
Near the boundary, the coordinate transformation between the Bañados and BTZ geometries reduces to 
\be \label{eq:asymptotic-form-coordinate-transformation}
\begin{split}
	e^{2i \f{{r_{h_\mathcal{O}}}}{\ell} \xi} \approx \frac{f(z)}{\overline{f}(\overline{z})},~e^{2 \f{{r_{h_\mathcal{O}}}}{\ell} \phi} \approx f(z)\, \overline{f}(\overline{z}),~r^2 \approx \frac{{r_{h_\mathcal{O}}}^2
 f(z) \overline{f}(\overline{z})}{y^2 f'(z) \overline{f}'(\overline{z})},
\end{split}
\ee
which yields
\begin{align}
	\xi &\approx \frac{1}{2i \f{{r_{h_\mathcal{O}}}}{\ell}} \log\left( \frac{f(z)}{\overline{f}(\overline{z})} \right), \quad
	\phi \approx \frac{1}{2 \f{{r_{h_\mathcal{O}}}}{\ell}} \log\left( f(z)\, \overline{f}(\overline{z}) \right).
\end{align}
The asymptotic geometry near the boundary, $r\approx \infty$, is given by
\begin{equation}
	ds_E^2 \underset{r\approx \infty}{\approx} \ell^2 \left(\frac{dr^2}{r^2} + \frac{r^2\, dz\, d\overline{z} \, f'(z)\, \overline{f}'(\overline{z})}{{r_{h_\mathcal{O}}}^2 f(z)\, \overline{f}(\overline{z})}\right).
\end{equation}
If we change the radial coordinate from $r$ to $\rho$ defined as 
\begin{equation}
	\rho = \frac{L}{2\pi}\left(\frac{\frac{\pi }{2}-\tan ^{-1}\left[ \left(  \frac{r^2 f'(z)\, \overline{f}'(\overline{z})}{{r_{h_\mathcal{O}}}^2 f(z)\, \overline{f}(\overline{z})}\right)^\f{1}{2}\right]}{\frac{\pi }{2}}\right)
	= \frac{L}{2\pi}\left(\frac{\frac{\pi }{2}-\tan ^{-1}\left[\left( r^2  \sqrt{\frac{ T(z)\, \overline{T}(\overline{z})}{h_{\mathcal{O}} \overline{h}_{\mathcal{O}}} }\right)^\f{1}{2}\right]}{\frac{\pi }{2}}\right),
\end{equation}
where \( T(z), \overline{T}(\overline{z}) \) are defined in \eqref{eq:energy-density-z}, then the asymptotic metric is given by the Poincaré AdS$_3$,
\be
ds^2_{E} \underset{\rho \approx 0}{\approx}
\ell^2 \left(\f{d\rho^{2}+dzd\overline{z}}{\rho^{2}}\right).
\ee
\subsection{Spacetime Dependence of Horizon}
\begin{figure}[htbp]
	\centering
	\subfloat[$\epsilon_1 = 10, {r_{h_\mathcal{O}}}=10, L= 10000, x= 2500, L=10000.$]{\includegraphics[width=.4\columnwidth]{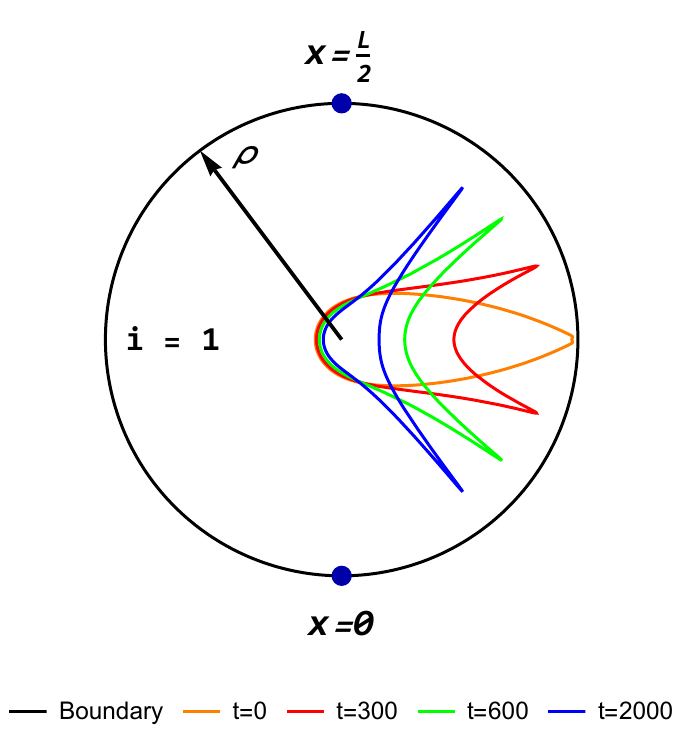}}\hspace{5pt}
	\subfloat[$\epsilon_1 = 10, {r_{h_\mathcal{O}}}=10, L= 10000, x= 2500, L=10000.$]{\includegraphics[width=.4\columnwidth]{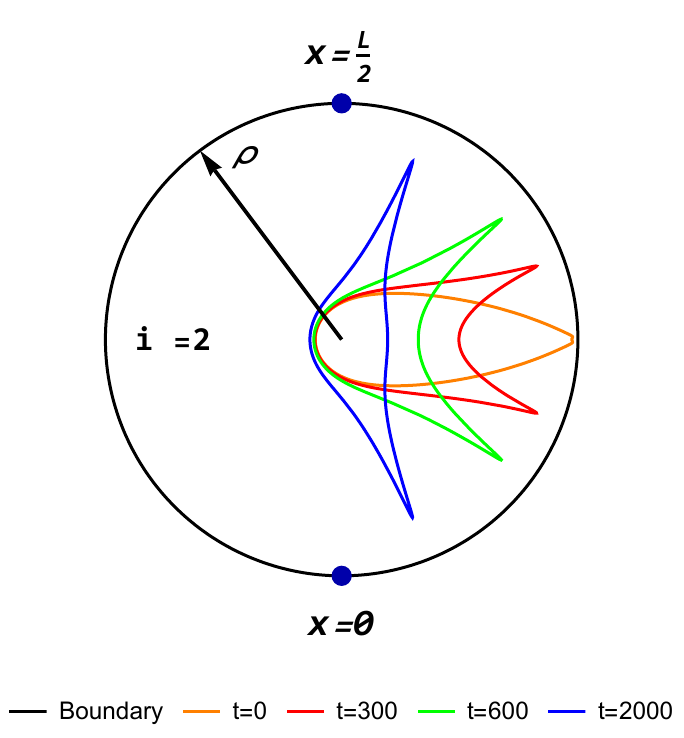}}\\
	\subfloat[$\epsilon_1 = 10, \epsilon_2 = 100, {r_{h_\mathcal{O}}}=10, L= 10000, x= 2500, L=10000.$]{\includegraphics[width=.4\columnwidth]{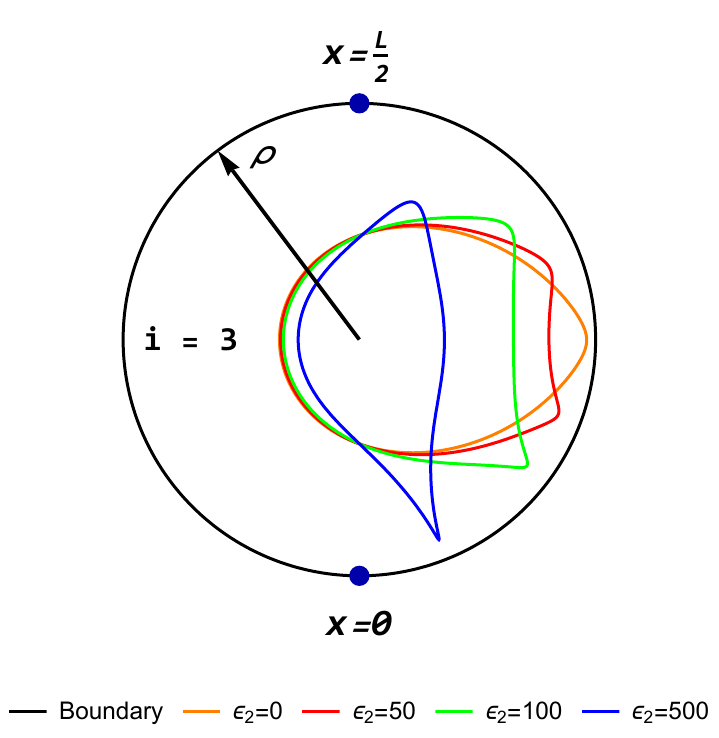}}\hspace{5pt}
	\subfloat[$\epsilon_1 = 10, \epsilon_2 = 100, {r_{h_\mathcal{O}}}=10, L= 10000, x= 2500, L=10000.$]{\includegraphics[width=.4\columnwidth]{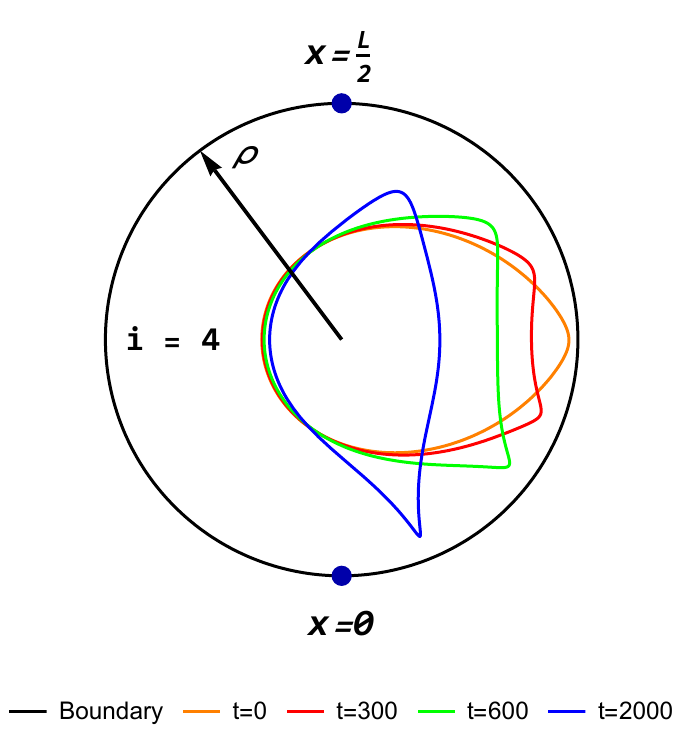}}
	\caption{The behaviors of the horizon for $i=1$ (I), $2$ (II), $3$ (III), and $4$ (IV) for several $t$. In this figure, the radial coordinate is $\rho$, and the curves in different colors represent the horizon position at various time slices. We can see that the peaks gradually approach to $x = 0$ and $x=\frac{L}{2}$.
    }\label{Fig:Horizon-2d-1}
\end{figure}

\begin{figure}[htbp]
	\centering
	\subfloat[$t=10000, \epsilon_1 = 10, {r_{h_\mathcal{O}}}=10, L= 10000, x= 2500, L=10000.$]{\includegraphics[width=.4\columnwidth]{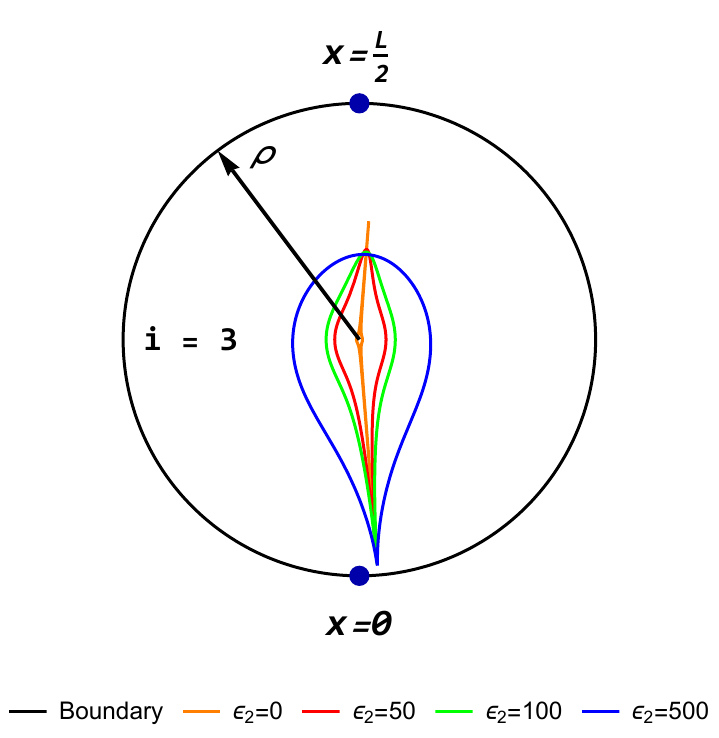}}\hspace{5pt}
	\subfloat[$t=10000, \epsilon_1 = 10, {r_{h_\mathcal{O}}}=10, L= 10000, x= 2500, L=10000.$]{\includegraphics[width=.4\columnwidth]{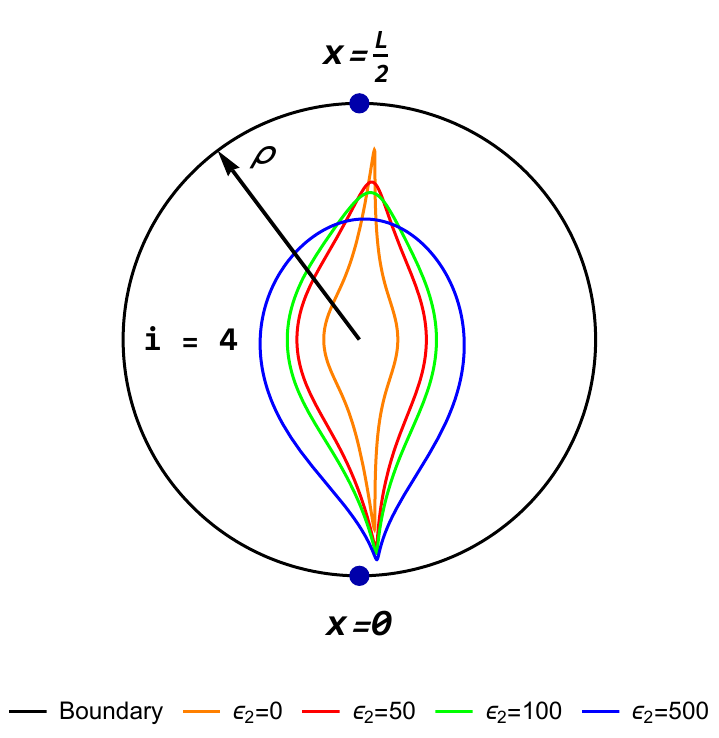}}\\
	\caption{The behavior of the horizon for $i=3$ (I) and $i=4$ (II) as a function of $\epsilon_{2}$ at a fixed time. As $\epsilon_{2}$ becomes larger and larger, the horizon becomes increasingly rounded except for the peak around $x=0$.
    }\label{Fig:Horizon-2d-2}
\end{figure}

Now, we will investigate the spacetime dependence of the horizon in the Poincaré coordinate.
To this end, we exploit the analytic continuation, $\tau =it$, and then map from $r=r_{h_\mathcal{O}}$ to 
\begin{equation} \label{eq:shape-of-horizon}
	\rho_h = \frac{L}{2\pi}\frac{\frac{\pi }{2}-\tan ^{-1}\left[ \left({r_{h_\mathcal{O}}}^2 \cdot \sqrt{\frac{T(z) \bar{T}(\bar{z})}{h_{\mathcal{O}} \bar{h}_{\mathcal{O}}}}\right)^\frac{1}{2}\right]}{\frac{\pi }{2}}.
\end{equation}
In Fig. \ref{Fig:Horizon-2d-1}, we show the shape of the black hole horizon in (\ref{eq:shape-of-horizon}) for $i=1,2,3,$ and $4$  as a function of $t$.
From this figure, we can observe the following behaviors of the black hole horizon.
For $i=1$ and $2$, the shape of the horizon initially looks elliptical. 
The black hole horizon at the point, where the local operator is inserted, is nearest to the boundary.
During the time evolution, the shape of the black hole inhomogeneously changes,  two peaks, approaching the boundary, emerge, and then they move closely to $X=0$ and $X=\f{L}{2}$.  For $i=3$ and $4$, the time dependence of the shape of the black hole horizon is similar to $i=1$ and $2$. The peaks of the black hole horizon, emerging under the time evolution, are more gentle than for $i=1$ and $2$.  
In Fig. \ref{Fig:Horizon-2d-2}, we further illustrate how the horizon profile at a fixed time depends on the deformation parameter $\epsilon_2$. As $\epsilon_2$ increases, the two emergent peaks of the horizon recede from the boundary, and the overall horizon shape becomes noticeably more rounded. In particular, for small $\epsilon_2$, one observes sharp, pronounced peaks that approach closer to $X=0$ and $X=\frac{L}{2}$, whereas larger values of $\epsilon_2$ suppress these sharp inhomogeneities, yielding a smoother, more nearly spherical horizon. One particularly notable observation is that the peak approaching $X=\frac{L}{2}$ lies farther from the boundary along the $\rho$ direction than the one approaching $X=0$.
This reflects that $e^{-\epsilon H_{SSD}}$ suppresses the excitation around $X=\f{L}{2}$ more than that around $X=0$. 
\if[0]
Thus, the position-dependent effective horizon radius is given by:
\begin{equation}
	r'_h(z, \overline{z}) = {r_{h_\mathcal{O}}} \left( \frac{T(z)\, \overline{T}(\overline{z})}{h_{\mathcal{O}} \overline{h}_{\mathcal{O}}} \right)^{1/4}.
\end{equation}

To visualize the gravitational response to different excitations, we define:
\begin{equation}
	\rho(X, t) = \frac{1}{r'_h(X, t)},
\end{equation}
and plot this function for all four local quench protocols \( i = 1,2,3,4 \). In the original \( (w, \overline{w}) \) coordinates, the boundary CFT is defined on a circle of radius \( L \). We may schematically illustrate the dynamical behavior of the black hole horizon as it evolves along this spatial circle. See Fig.~\ref{Fig:Horizon-2d-1} for a depiction of this process.
\fi

 \section{Discussion and future directions \label{sec:discussion-and-future-directions}}
We close this paper with the discuss on our main findings of this paper and future directions. 
Our findings in this paper are:
\begin{itemize}
   \item[1.]The time ordering of the Euclidean and Lorentzian time evolution operators acting on the local operator influences the entanglement dynamics. Especially, this time order influences the late-time behavior of the entanglement entropy and mutual information: for $i=1$, they can logarithmically grow in time; for $i=2$, they can saturate to the constant. 
   
    \item [2.] If in $i=3$ and $4$, the damping factor suppresses one of the quasiparticles forming an entanglement pair, then the bipartite and non-local correlation may be suppressed.
    This supports the idea that bipartite entanglement and non-local correlation induced by the local operator can be interpreted as the entanglement between the quasiparticles induced by the local operator. 
    \item[3.] In (\ref{eq:entanglement-entropy-and-energy-density}), we established the relation between the entanglement entropy and the chiral and anti-chiral energy-momentum densities.  By exploiting this formula, we found that the time ordering for $i=1$ induces the excitation widely distributed in the system, so that this excitation results in the logarithmic growth of the entanglement entropy and mutual information. 
   In contrast, the time ordering for $i=2$ induces the local excitations, so that the entanglement entropy and mutual information saturate to a certain value.
\end{itemize}

The first and third findings suggest that the time ordering of the time evolutions makes a significant difference between the behavior of the energy-momentum densities for $i=1$ and $i=2$, and then the difference of the energy-momentum densities at the endpoints of the subsystem makes the late time behavior of $S_{A,i=1}$ differ from that of $S_{A,i=2}$. 
For $i=1$, the energy-momentum densities at the endpoints of the subsystem behave in time as $t^{-4}$, and then decay to the vacuum ones, while for $i=2$, they saturate to the non-vacuum values.
This suggests that this remarkably different behavior of $S_{A,i=1}$ and $S_{A,i=2}$ may be due to the difference in the energy distribution. 
 Also, since $S_{A;i=1}$ could become infinite in the very late-time regime even for the finite subsystem, the time ordering for $i=2$ seems to be a natural choice.
As pointed out in Refs. \cite{2025arXiv250501508L,2020arXiv200810322L}, the Euclidean time evolution may be thought of as the measurement.
Therefore, we may be able to make a bridge between the time ordering effect and the measurement-induced phase transition, a remarkable dynamical phenomenon induced by the measurement \cite{2020PhRvR...2c3347R,2021Quant...5..528B,2019PhRvB.100m4306L,2019PhRvX...9c1009S,2019PhRvB..99v4307C,2019PhRvB.100f4204S,2020PhRvX..10d1020G,2020PhRvB.101f0301Z,2021Quant...5..528B,2021PhRvB.103v4210T,2021PhRvL.126q0503G,2023ScPP...14..138L,2023PhRvL.130w0401G,2024PhRvR...6a3131S,2024PhRvR...6b3081Z,2023PhRvX..13b1026G,2023PhRvB.108p5120Y,2023PhRvX..13d1042M,2024ScPP...17...20M,2020PhRvL.125c0505C}.

In the second main finding, by exploiting the entanglement entropy and mutual information, we investigated if the bipartite entanglement and non-local correlation is due to the entanglement between quasiparticles.
It is also intriguing to explore if the quantum correlation is due to quasiparticles by exploiting the logarithmic negativity \cite{Vidal:2002zz,Calabrese:2012ew}.

The third main finding is the fundamental relation between the entanglement entropy and the energy-momentum densities in $2$d holographic CFTs. 
As in this paper, by exploiting the map from the Bañados geometry to the geometry under consideration, we can explore the relation between other entanglement measures, such as the reflected entropy \cite{2019arXiv190500577D}, and energy-momentum densities.
It is intriguing to explore the laws, like the first law of thermodynamics, connecting the increases in entanglement measures and internal energy from this formula.
We leave it as a future direction.
\section*{Acknowledgments}
We thank useful discussions with Tadashi Takayanagi, Chen Bai, Farzad Omidi, and Yuxuan Zhang. 
M.N.~is supported by funds from the University of Chinese Academy of Sciences (UCAS), funds from the Kavli
Institute for Theoretical Sciences (KITS).

\appendix
\section{Deformed Hamiltonians and Quasiparticle Picture in 2D CFT}\label{ch:apa}
In this appendix, we will discuss the geometric origin of deformed Hamiltonians in two-dimensional conformal field theory (2D CFT), highlighting their relation to conformal Killing vectors and topological surface operators. We will also explain how the 2-SSD Hamiltonian leads to asymmetric quasiparticle propagation. Some steps follow \cite{10.21468/SciPostPhys.10.2.049} and are included here for completeness.
\subsection{Deformed Hamiltonians and Conformal Symmetry}

Consider the Hamiltonian:
\begin{equation}
	H_v = \int_0^L \frac{dx}{2\pi} \left[ v(x) T(x) + \bar{v}(x) \bar{T}(x) \right],
\end{equation}
where \( v(x) \), \( \bar{v}(x) \) are smooth, periodic deformation functions. The standard CFT Hamiltonian corresponds to \( v = \bar{v} = 1 \).

In general dimension \( d \), conformal symmetries are generated by Killing vectors \( \xi^\mu \), with corresponding surface charges:
\begin{equation}
	Q_\xi(\Sigma) = -\int_\Sigma d\sigma^\mu\, \xi^\nu T_{\mu\nu},
\end{equation}
which are topological under the small deformations of the codimension-one surface \( \Sigma \).

In \( d=2 \), the conformal Killing equation admits infinitely many solutions:
\begin{equation} \label{eq:killing-vector}
	\partial_{\bar{z}} \xi^z = 0, \quad \partial_z \xi^{\bar{z}} = 0,
\end{equation}
with complex coordinates defined as:
\begin{equation}
	z = x^0 + i x^1, \quad \bar{z} = x^0 - i x^1, \quad 
	\partial_z = \frac{1}{2}(\partial_0 - i \partial_1), \quad 
	\partial_{\bar{z}} = \frac{1}{2}(\partial_0 + i \partial_1).
\end{equation}
The solutions of (\ref{eq:killing-vector}) take the form \( \xi^z = \xi(z) \), \( \xi^{\bar{z}} = \bar{\xi}(\bar{z}) \), with arbitrary holomorphic and anti-holomorphic functions. The associated charges are:
\begin{equation}
	Q_\xi = \oint_\Sigma \frac{1}{2\pi i} \left[ dz\, \xi(z) T(z) - d\bar{z}\, \bar{\xi}(\bar{z}) \bar{T}(\bar{z}) \right].
\end{equation}

Under Wick rotation \( x^0 \rightarrow i x^0 \), we parametrize:
\begin{equation}
	z = i(x^0 + x^1), \quad \bar{z} = i(x^0 - x^1).
\end{equation}
Fixing \( x^0 = 0 \), the chiral operator evolution under \( U = e^{-i t Q_\xi} \) corresponds to:
\begin{equation}
	\frac{d x_t}{dt} = v(x_t), \quad v(x) := \xi(z)|_{z = i x},
\end{equation}
and similarly for the anti-chiral case:
\begin{equation}
	\frac{d x_t}{dt} = -\bar{v}(x_t), \quad \bar{v}(x) := \bar{\xi}(\bar{z})|_{\bar{z} = -i x}.
\end{equation}

For a finite system with length of \( L \), one maps the spatial line to a circle:
\begin{equation}
	z = e^{\frac{2\pi}{L} w}, \quad \bar{z} = e^{\frac{2\pi}{L} \bar{w}}, \quad
	w = \tau + i x, \quad \bar{w} = \tau - i x,
\end{equation}
and expands:
\begin{equation}
	\xi(z) = z \sum_n \tilde{v}_n z^n, \quad 
	\bar{\xi}(\bar{z}) = \bar{z} \sum_n \tilde{\bar{v}}_n \bar{z}^n.
\end{equation}
The corresponding surface charge becomes :
\begin{equation}\label{eq:surface-charge}
	Q_\xi = \frac{L}{2\pi} \int_0^L \frac{dx}{2\pi} \left[ v(x) T(x) + \bar{v}(x) \bar{T}(x) \right] + \frac{v_0 c + \bar{v}_0 \bar{c}}{24},
\end{equation}
where we have used the transformation rule for the chiral energy-momentum tensor
\begin{equation}
	\begin{aligned}
		T(w)=\left(\frac{\partial z}{\partial w}\right)^2 T(z)+\frac{c}{12} \operatorname{Sch}(z, w)=\left(\frac{2 \pi}{L}\right)^2\left[e^{\frac{4 \pi}{L} w} T(z)-\frac{c}{24}\right],
	\end{aligned}
\end{equation}
and $v(x)$ and $\overline{v}(x)$ are given by
\begin{equation}
	v(x) = \sum_n \tilde{v}_n e^{i \frac{2\pi n}{L} x}, \quad 
	\bar{v}(x) = \sum_n \tilde{\bar{v}}_n e^{-i \frac{2\pi n}{L} x}.
\end{equation}
We identify the integral term in \eqref{eq:surface-charge} as the deformed Hamiltonian on the circle:
\begin{equation}
	H_v = \int_0^L \frac{dx}{2\pi} \left[ v(x) T(x) + \bar{v}(x) \bar{T}(x) \right].
\end{equation}

The time evolution of a chiral operator \( O(w) \) under \( H_v \) obeys:
\begin{equation}
	\frac{d z_t}{dt} = i \frac{2\pi}{L} \xi(z_t) \quad \Rightarrow \quad \frac{d w_t}{dt} = i v(-i w_t),
\end{equation}
which reduces on the \(\tau = 0\) slice (\( w = i x \)) to:
\begin{equation}
	\frac{d x_t}{dt} = v(x_t),
\end{equation}
and similarly \( \frac{d x_t}{dt} = -\bar{v}(x_t) \) for anti-chiral operators.

\subsection{Quasiparticle Dynamic under 2-SSD Hamiltonian}

Now, we will explain how the quasiparticles propagate to spatial locations under the time evolution induced by 2-SSD Hamiltonian.
This Hamiltonian is given by
\begin{equation}
	H_{2\text{-SSD}} = 2 \int_0^L dx\, \sin^2\left(\frac{2\pi x}{L}\right) T_{00}(x),
\end{equation}
which induces position-dependent group velocities,
\begin{equation}
	\left|v(x)\right|  = 2 \sin^2\left(\frac{2\pi x}{L}\right),
\end{equation}
vanishing at \( x = 0 \) and \( x = L/2 \), and peaking at \( x = L/4 \) and \( x = 3L/4 \).
Consider a pair of entangled quasiparticles generated around \( x = L/4 \), with one moving toward \( x = 0 \) and the other toward \( x = L/2 \). Since the velocities around $x=0$ or $x=\f{L}{2}$ exponentially small, a quasiparticle moves to $x=0$, while another moves to $\f{L}{2}$. Consequently, one quasiparticle accumulates around $x=0$, while the other does around $x=\f{L}{2}$.

Let us evaluate the time \( t \) that it takes for a quasiparticle with this velocity to travel from position \( x_0 \in (0, L/2) \) to either endpoint under this velocity profile. The exact expression of $t$ is
\begin{equation}
	t(x_0 \to x) = \left|\int_{x_0}^{x} \frac{dy}{2 \sin^2\left( \frac{2\pi y}{L} \right)}\right| = \left|\frac{L}{4\pi} \left[ \cot\left( \frac{2\pi x_0}{L} \right) - \cot\left( \frac{2\pi x}{L} \right) \right]\right|,
\end{equation}

As \( x \to 0 \) or \( x \to L/2 \), the argument of the cotangent function approaches a multiple of \( \pi \), and thus
\begin{equation}
	\cot\left( \frac{2\pi x}{L} \right) \to \pm \infty,
\end{equation}
implying that \( t \to \infty \). Therefore, quasiparticles never reach \( x = 0 \) or \( x = L/2 \) in finite time, but only asymptotically approach them.

This result explains  why one quasiparticle localizes around \( x = 0 \), while the other does around \( x = L/2 \). 

\section{Entanglement entropy with various $h_{\mathcal{O}}$} \label{ch:special}

\begin{figure}[htbp!]
	\centering
	\subfloat[$\delta=1, \epsilon_1 = 5, \epsilon_2 = 0, L= 10000, x= 2500, X_2=3000, X_1=6000.$]{\includegraphics[width=.4\columnwidth]{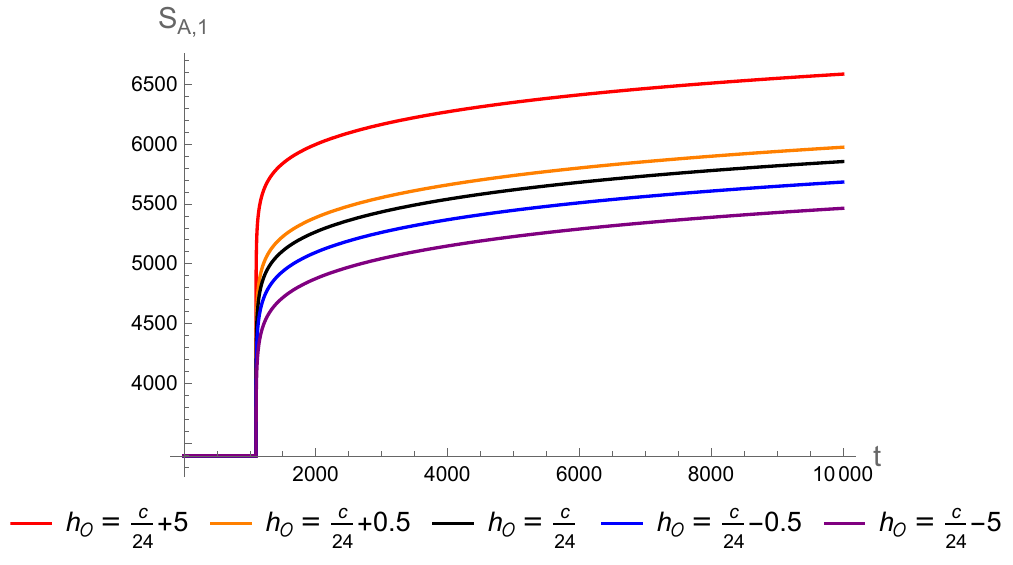}}\hspace{5pt}
	\subfloat[$\delta=1, \epsilon_1 = 5, \epsilon_2 = 0, L= 10000, x= 2500, X_2=3000, X_1=6000.$]{\includegraphics[width=.4\columnwidth]{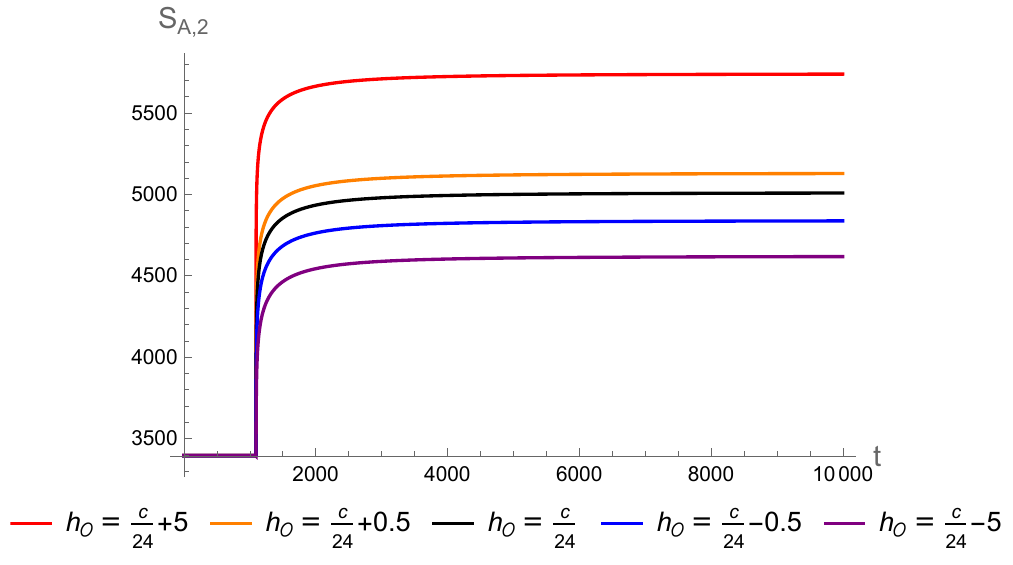}}\\
		\subfloat[$\delta=1, \epsilon_1 =5, \epsilon_2 = 40, L= 10000, x= 2500, X_2=3000, X_1=6000.$]{\includegraphics[width=.4\columnwidth]{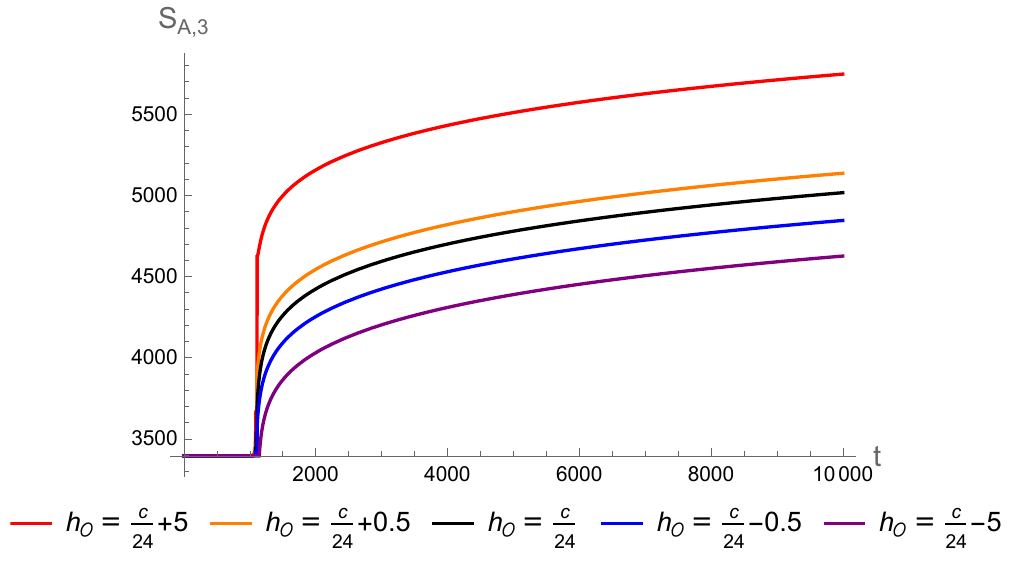}}\hspace{5pt}
	\subfloat[$\delta=1, \epsilon_1 = 5, \epsilon_2 = 40, L= 10000, x= 2500, X_2=3000, X_1=6000.$]{\includegraphics[width=.4\columnwidth]{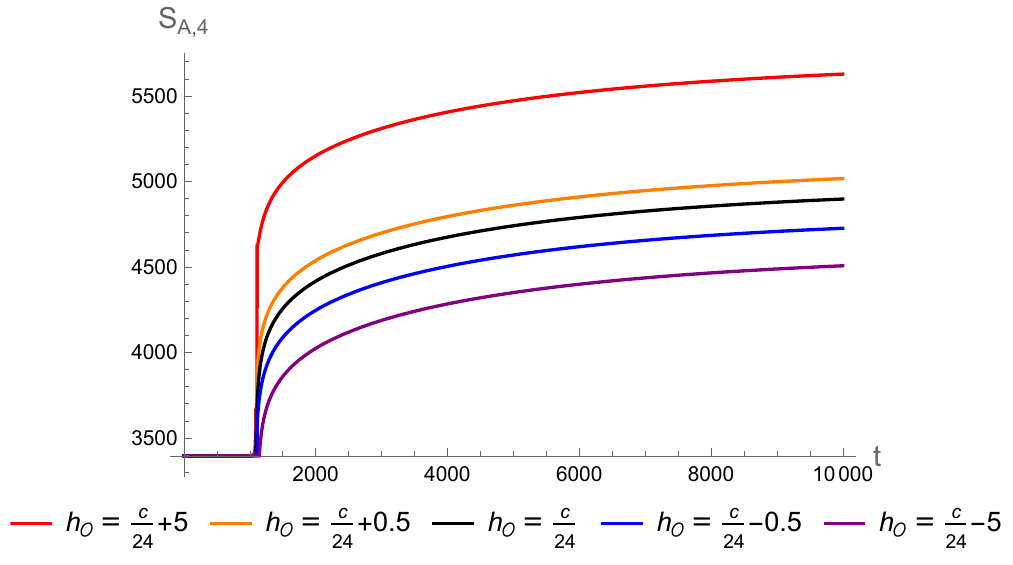}}\\
	\caption{ Time evolution of the entanglement entropy $S_{A,i}(t)$ for four different excited states labeled by $i=1,2,3$, and $4$, corresponding to the Heisenberg operators $\mathcal{O}_{H,i}$ discussed in the main text. In each panel, the different curves represent the time dependence of $S_{A,i}$ for the distinct conformal dimensions $h_{\mathcal{O}}$ of the inserted local operator. The smooth interpolation among the curves near $h_{\mathcal{O}} = \frac{c}{24}$ suggests continuity in the dynamical behavior of the entanglement entropy across these values.}\label{Fig:different-h-1}
\end{figure}

When \( h_{\mathcal{O}} = \frac{c}{24} \), we have $T(z)=\frac{c}{24} \left( \frac{1}{z - z_{\epsilon}^{\text{new}, a}} - \frac{1}{z - z_{-\epsilon}^{\text{new}, a}} \right)^2$. Thus, Eq.~\eqref{eq:ODE-f-T} admits the general solution \( f(z) = c_3 -\frac{c_2}{(z^{\text{new},a}_{\epsilon} - z^{\text{new},a}_{-\epsilon}) \left[ (z^{\text{new},a}_{\epsilon} - z^{\text{new},a}_{-\epsilon}) c_1 + \log(z - z^{\text{new},a}_{\epsilon}) - \log(z - z^{\text{new},a}_{-\epsilon}) \right]}
 \). 
For simplicity, we fix  \( f(z) =\frac{1}{\log(z - z^{\text{new},a}_{\epsilon}) - \log(z - z^{\text{new},a}_{-\epsilon}) } \) and \( \bar{f}(\bar{z}) =\frac{1}{\log(\bar{z} - \bar{z}^{\text{new},a}_{\epsilon}) - \log(\bar{z} - \bar{z}^{\text{new},a}_{-\epsilon}) } \) without loss of generality.
Thus, from our definition of $\mathcal{T}(z)$, we have
\begin{equation}
\mathcal{T}(z)=\frac{z^{\text{new},a}_{-\epsilon}-z^{\text{new},a}_{\epsilon}}{(z-z^{\text{new},a}_{\epsilon}) (z-z^{\text{new},a}_{-\epsilon}) (\log (z-z^{\text{new},a}_{\epsilon})-\log (z-z^{\text{new},a}_{-\epsilon}))},
\end{equation}
with an analogous expression for the anti-holomorphic part as
\begin{equation}
	\mathcal{\overline{T}}(\bar{z})=\frac{\bar{z}^{\text{new},a}_{-\epsilon}-\bar{z}^{\text{new},a}_{\epsilon}}{(\bar{z}-\bar{z}^{\text{new},a}_{\epsilon}) (\bar{z}-\bar{z}^{\text{new},a}_{-\epsilon}) (\log (\bar{z}-\bar{z}^{\text{new},a}_{\epsilon})-\log (\bar{z}-\bar{z}^{\text{new},a}_{-\epsilon}))}
\end{equation}
Thus, the entanglement entropy can be obtained from  Eq.\eqref{eq:general-entanglement-entropy}.
Consider the subsystem $A$ containing only a single fixed point at $x=0$.
Then, we show the time dependence of $S_{A,i=1\sim 4}$ for various $h_{\mathcal{O}}$ in Fig. \ref{Fig:different-h-1}.

\newpage
\bibliographystyle{ieeetr}
\bibliography{reference.bib}
\end{document}